\documentclass[a4paper,11pt]{article}
\pdfoutput=1 

\usepackage{jheppub} 

\usepackage[T1]{fontenc} 
\usepackage{braket}
\usepackage[dvipsnames]{xcolor}
\usepackage{comment}
\usepackage{float}
\usepackage{caption}
\usepackage{subcaption}
\usepackage{bbold}  
\usepackage{hyperref}
\newcommand{\al}{{\alpha}}
\newcommand{\bt}{{\beta}}
\newcommand{\de}{{\delta}}
\newcommand{\ep}{{\epsilon}}

\newcommand{\te}{{\theta}}
\newcommand{\sig}{{\sigma}}

\newcommand{\pd}{{\partial}}

\usepackage{amsmath}
\DeclareMathOperator{\arcsinh}{arcsinh}

\title{\boldmath Thin-wall vacuum decay in the presence of a compact dimension}

\author[a,b]{Ignatios Antoniadis}
\author[a]{Daniele Bielli}
\author[a]{Auttakit Chatrabhuti}
\author[a]{Hiroshi Isono}

\affiliation[a]{High Energy Physics Research Unit, Faculty of Science,
Chulalongkorn University,
\\
Bangkok 10330, Thailand}
\affiliation[b]{Laboratoire de Physique Th\'eorique et Hautes Energies - LPTHE
\\
Sorbonne Universit\'e, CNRS, 4 Place Jussieu, 75005 Paris, France}

\emailAdd{antoniad@lpthe.jussieu.fr, d.bielli4@gmail.com, auttakit.c@chula.ac.th, hiroshi.isono81@gmail.com}

\abstract{We study the problem of false vacuum decay in arbitrary dimensions, in the presence of gravity, and compute the transition probability within the thin-wall approximation, generalising the results of Coleman and de Luccia. In the particular case of one compact dimension, we present explicit formulae for the Euclidean Bounce configuration that drives the transition from a de Sitter to Minkowski or from a Minkowski to anti-de Sitter vacua.}

\begin{document} 
\maketitle
\flushbottom

\section{Introduction}
The false vacuum decay is one of the important phenomena in quantum theories, already at the level of quantum mechanics, with interesting applications to a variety of areas, such as phase transitions, non perturbative dynamics of quantum field theories and cosmology. A standard approach to study the decay is by saturating the Euclidean path integral with a classical configuration of finite action that extrapolates between the false and true vacuum corresponding to a bubble of the true vacuum inside the false vacuum, the so-called Bounce~\cite{Coleman:1977py, Coleman:1980aw, Hawking:1981fz}. The study of the Bounce in a scalar quantum field theory was initiated long ago in the thin-wall approximation and the decay width was computed by extremising the action with respect to the position of the wall~\cite{Coleman:1977py}. The analysis was also generalised in the presence of gravity and explicit expressions were given for the transitions from a de Sitter (dS) to Minkowski and from a Minkowski to anti-de Sitter (AdS) vacuum~\cite{Coleman:1980aw}. It was later realised that the thin-wall approximation can be interpreted as a 3-dimensional Euclidean world-volume of a brane separating the false and the true vacuum. The extremisation condition can be replaced by imposing the Israel matching conditions on the brane~\cite{Israel:1966rt}, leading to the same result~\cite{Garfinkle:1989mv,Chamblin:1999ya} (for a recent review see e.g.~\cite{Gregory:2023eos} and references therein).

The nucleation of a true vacuum bubble inside the false vacuum is a quantum tunneling event with a probability of occurrence per unit time given by the decay width per unit volume $\Gamma/V$, which in the semiclassical limit admits an expansion of the form
\begin{equation}
\frac{\Gamma}{V} = A e^{-\frac{B}{\hbar}} (1+\mathcal{O}(\hbar))\,,
\end{equation} 
where $B$ is the action of the Bounce which is usually large and gives the dominant contribution to the transition probability. The thin-wall approximation is valid in the case where the two vacua are almost degenerate with a high potential barrier. The computation beyond the thin-wall approximation can in principle be done numerically, based for instance on the under-shooting/over-shooting method~\cite{Coleman:1977py}, or semi-analytically depending on the model~\cite{Lee:1985uv, Jensen:1983ac, Espinosa:2018hue, Espinosa:2018voj}. In the other limit of a very shallow potential around the two minima, the Coleman-de Luccia (CdL) instanton becomes subdominant and the transition is mainly driven by a classical field trajectory over the barrier~\cite{Hawking:1981fz, Balek:2003uu}. 

The generalisation of the false vacuum decay in various dimensions $d\ne 4$ presents obviously an interest in critical phenomena for $d<4$, while for $d>4$ is also motivated by the role of extra dimensions in physics beyond the Standard Model, cosmology and string theory. In the non compact case, the generalisation is in principle straightforward and several results have already been obtained recently in this direction~\cite{Espinosa:2022jlx, Matteini:2024xvg, Espinosa:2024ufg}. 

In this work, we study the false vacuum decay in the presence of compact dimensions, restricting to the simplest illustrative case of a single one\footnote{For interesting related work involving the presence of non-dynamical compact directions see for example~\cite{Kleban:2011cs, Apruzzi:2019ecr, Apruzzi:2021nle}.
For non-Euclidean approaches to the vacuum decay problem see for example \cite{Garcia-Compean:2021syl}. }. 
The main complication comes from the symmetry of the Bounce which cannot be spherical $O(d)$ but $O(d-1)\times U(1)$, describing a bubble geometry of a $S^{d-2}$ sphere times a circle $S^1$ with their respective radii depending on the radial coordinate.
This is parameterised by two functions of the radial coordinate, associated to the radius of $S^{d-2}$ and to the radius of the compact dimension. It turns out that the extremisation procedure of the Bounce as in CdL becomes ambiguous while the Israel matching conditions are well defined and lead to simple explicit expressions for the Bounce and for the transition probabilities in the cases of dS to Minkowski and Minkowski to AdS vacua.

Note that in Lorentzian metric with flat spatial sections, there is a maximally symmetric $d$-dimensional solution of the vacuum equations of motion corresponding to a dS or AdS spacetime with one compact spatial dimension. The Euclidean version of these solutions does not have however finite volume and thus the Euclidean action is not finite.

The outline of our paper is the following. In Section~2, we present an overview of the false vacuum decay in arbitrary non compact dimensions driven by a Bounce with $O(d)$ spherical symmetry and compute its action in the thin-wall approximation using both the CdL approach (subsection 2.1) and the Israel matching conditions (subsection 2.2). In Section~3, we extend the analysis in the presence of one compact extra dimension. In particular, we compute the instanton geometry (subsection 3.1), we discuss the difficulties of the CdL approach (subsection 3.2) and proceed with the derivation of the Israel matching conditions which determine uniquely the bubble solution of the equations of motion (subsection 3.3). In Section~4, we compute the decay width of the false vacuum specialising to the transitions of dS to Minkowski (subsection 4.1) and Minkowski to AdS (subsection 4.2) vacua. Finally, Section~5 contains our conclusions and outlook.

\section{Vacuum Decay in Theories with $O(d)$ Symmetry}\label{sec:theories-with-O(d)-symmetry}
In this section we revise the evaluation of vacuum transition amplitudes, in the thin-wall approximation, for the theory of a scalar field $\phi$ with potential $V(\phi)$ on a $d$-dimensional background enjoying $O(d)$ symmetry. The starting point is the following Euclidean action 
\begin{equation}\label{Euclidean-action}
S_{\mathrm E}[\phi] = \int \mathrm{d}^{d}x \sqrt{\tilde{g}} \, \biggl( \frac{1}{2}\tilde{g}^{\mu\nu}\nabla_{\mu}\phi\nabla_{\nu}\phi + V(\phi) -\frac{1}{2\kappa_{d}}R[\tilde{g}] \biggr) + S_{\mathrm{GHY}} \, \, ,
\end{equation}
where $S_{\mathrm{GHY}}$ represents the Gibbons-Hawking-York boundary action, which will be given below shortly.
The background metric $\tilde{g}$, with scalar curvature $R[\tilde{g}]$, depends on a single unknown function of the radial direction. This can be written in the following form
\begin{equation}\label{gtilde-metric-CdL}
\tilde{g} = \mathrm{d}\xi^2 + \rho^2(\xi) \mathrm{d}\Omega_{d-1}^2 
= \mathrm{d}\xi^2+\rho^2(\xi)g_{ij}\mathrm{d}y^i\mathrm{d}y^j \, ,
\end{equation}
where $\mathrm{d}\Omega_{d-1}^2=g_{ij}\mathrm{d}y^i\mathrm{d}y^j$ represents the surface element of a $(d-1)$-dimensional sphere of radius $H^{-1}$. In the following we shall denote derivatives with respect to $\xi$ as
\begin{equation}
\begin{aligned}
\dot{\rho}(\xi)\equiv \frac{\mathrm{d}\rho(\xi)}{\mathrm{d}\xi} \, ,
\qquad
\ddot{\rho}(\xi)\equiv \frac{\mathrm{d}^2\rho(\xi)}{\mathrm{d}\xi^2} \, .  
\end{aligned}
\end{equation}
In terms of them, the Gibbons-Hawking-York action $S_{\mathrm{GHY}}$ is given by
\begin{align}
S_{\mathrm{GHY}}=-\frac{1}{\kappa_d}\int\!\mathrm{d}^dx\,\frac{\mathrm{d}}{\mathrm{d}\xi}(\sqrt{g}K) \, ,
\end{align}
with $K$ the extrinsic curvature associated with the spherical hypersurface at each $\xi$,
\begin{align}
K=\frac{d-1}{2}\rho^{-2}\pd_\xi\rho^2|_{\xi=\bar\xi}=\frac{d-1}{2}\frac{\dot\rho(\bar\xi)}{\rho(\bar\xi)} \, .
\end{align}
The scalar curvature term in the Euclidean action \eqref{Euclidean-action} with the metric \eqref{gtilde-metric-CdL} gives rise a term with $\ddot\rho$. Integrating this term by part generates boundary terms which can be cancelled by $S_{\mathrm{GHY}}$, thus leaving us with
\begin{equation}\label{CdL-action}
\begin{aligned}
S_{E}[\phi] 
&= \Sigma(S^{d-1})\int_{0}^{\xi_{max}} \mathrm{d}\xi \, \rho^{d-1} \biggl( \frac{1}{2}\dot{\phi}^2 + V -\frac{(d-1)(d-2)}{2\kappa_{d}} \rho^{-2} ( H^2 + \dot{\rho}^2 ) \biggr) \, .
\end{aligned}
\end{equation}
where $\Sigma(S^{d-1})$ is the area of the $(d-1)$-dimensional unit sphere, given by
\begin{equation}\label{Sigdef}
\Sigma(S^{d-1})=\frac{2\pi^{\tfrac{d}{2}}}{\Gamma(\tfrac{d}{2})} \, .
\end{equation}
The Einstein equations, which are given in general by\footnote{It is convenient here and below to use the following identities:
\begin{align}
&R[\tilde g]_{\xi\xi}=-(d-1)\frac{\ddot\rho}{\rho}, \quad
R[\tilde g]_{ij}=R[g]_{ij}-[(d-2)\dot\rho^2+\rho\ddot\rho]g_{ij} \,, \\
&R[\tilde g]=\frac{R[g]}{\rho^2}-\left[(d-1)(d-2)\frac{\dot\rho^2}{\rho^2}+2(d-1)\frac{\ddot\rho}{\rho}\right] \,, \\
&R[\tilde g]_{ij}=R[g]_{ij}-[(d-3)\dot\rho^2+\dot\rho^2+\rho\ddot\rho]g_{ij}
=R[g]_{ij}-(d-3)\dot\rho^2h_{ij}-\frac{1}{2}\frac{\mathrm{d}^2\rho^2}{\mathrm{d}\xi^2}g_{ij} \,,
\end{align}
together with $R[g]=(d-1)(d-2)H^2$ and $R_{ij}[g] = g_{ij}(d-2)H^2$ for the metric $g$ for the $(d-1)$-dimensional sphere of radius $H^{-1}$.}
\begin{equation}
R_{\mu\nu}[\tilde g]-\frac{1}{2}\tilde g_{\mu\nu}R[\tilde g]=-\kappa_d\left(-\nabla_{\mu} \phi \nabla_{\nu}\phi + \frac{1}{2}\tilde{g}_{\mu\nu} \tilde{g}^{\alpha\beta} \nabla_{\alpha}\phi \nabla_{\beta}\phi+\tilde{g}_{\mu\nu}V\right) 
\end{equation}
read
\begin{align}
&\dot\rho^2=H^2+\frac{2\kappa_d\rho^2}{(d-1)(d-2)}\left(\frac{1}{2}\dot\phi^2-V(\phi)\right)\,, \label{Eeqd1} \\
&\dot\rho^2=H^2-\frac{2}{d-3}\rho\ddot\rho-\frac{2\kappa_d}{(d-2)(d-3)}\rho^2\left(\frac{1}{2}\dot\phi^2+V(\phi)\right) \,, \label{Eeqd2}
\end{align}
where the first is the Friedmann equation coming from the $(\xi\xi)$ component and the second comes from the spacial components. Note that the second equation \eqref{Eeqd2} coincides with the equation of motion (EOM) obtained by the variation of $S_{\textrm{E}}$ \eqref{CdL-action} with respect to $\rho$.

Let us solve them for a constant configuration of $\phi$ for which $\dot\phi=0$ and the potential takes a constant value $V_c$. Removing $\dot\rho^2$ from \eqref{Eeqd2} by using \eqref{Eeqd1}, we obtain
\begin{align}\label{rhoeqd}
\ddot\rho=-\frac{2\kappa_dV_c}{(d-1)(d-2)}\rho\,.
\end{align}
Solving this and substituting the solution into the Friedmann equation \eqref{Eeqd1}, we find the vacuum solution that can be classified based on the sign of $V_c$:
\begin{align}
V_c>0: &\quad \rho(\xi) = H\sqrt{\frac{(d-1)(d-2)}{2\kappa_{d}V_{c}}}\sin \biggl(\xi \sqrt{\frac{2\kappa_{d}V_{c}}{(d-1)(d-2)}} + \xi_0 \biggr) \, , \label{Vacuum-solution-CdL1} \\
V_c=0: &\quad \rho(\xi) = H(\xi-\xi_0) \, , \label{Vacuum-solution-CdL2} \\
V_c<0: &\quad \rho(\xi) = H\sqrt{\frac{(d-1)(d-2)}{2\kappa_{d}|V_{c}|}}\sinh \biggl(\xi \sqrt{\frac{2\kappa_{d}|V_{c}|}{(d-1)(d-2)}} + \xi_0 \biggr) \, .\label{Vacuum-solution-CdL3}
\end{align}
with $\xi_0$ an integration constant.

As just shown, it was quite easy to solve the equations of motion in this maximally symmetric case, but the method above, which derives an equation of harmonic oscillator type, does not apply to the case with one compact dimension. Notice however that one can derive from \eqref{Eeqd2} an equation that looks like a Friedmann equation \eqref{Eeqd1}. This goes as follows: first, we derive an equation for $\ddot\rho$ by taking a $\xi$-derivative of \eqref{Eeqd1} and substitute the equation of motion for $\phi$, which is $\ddot\phi+(d-1)(\dot\rho/\rho)\dot\phi-V'=0$. Next, we decompose $\rho\ddot\rho$ in \eqref{Eeqd2} as $((d-3)+(5-d))\rho\ddot\rho/2$, and substitute the expression for $\ddot\rho$ just obtained into $(5-d)\rho\ddot\rho/2$. We then find
\begin{align}
(d-2)(d-1)[H^2-(\dot\rho^2+\rho\ddot\rho)]=(d-3)\kappa_d\rho^2\dot\phi^2+4\rho^2\kappa_dV\,.
\end{align}
Let us now assume $\phi$ and $V=V_{c}$ are constant. 
Multiplying by $\rho\dot\rho$ gives a total derivative,
\begin{align}
\frac{\mathrm{d}}{\mathrm{d}\xi}\left[H^2\rho^2-\rho^2\dot\rho^2-\frac{2\kappa_dV_{c}}{(d-1)(d-2)}\rho^4\right]=0\,.
\end{align}
This is easily solved with an integration constant $\bt$ by the following Friedmann-like equation
\begin{align}\label{Friedmann-like}
\dot\rho^2=\frac{\bt}{\rho^2}+H^2-\frac{2\kappa_d\rho^2V_{c}}{(d-1)(d-2)} \,.
\end{align}
Comparing this with the Friedmann equation \eqref{Eeqd1} at constant $\phi$ and $V$ forces $\bt=0$.

In the present case, the Friedmann-like equation \eqref{Friedmann-like} happened to be identical, up to $\bt$, to the Friedmann equation \eqref{Eeqd1}. However, as will be shown later, in the case with a compact dimension, a Friedmann-like equation that is derived from a spatial component of Einstein equations in a similar manner is clearly different from the Friedmann equation and is indeed solvable for $\rho$.

\subsection{The approach of Coleman and de Luccia (CdL)}\label{subsec:CdL}
In this subsection we proceed in evaluating the bounce configuration which drives the vacuum transition by following the reasoning presented in \cite{Coleman:1980aw}. In particular, without exploiting the solution \eqref{Vacuum-solution-CdL1}--\eqref{Vacuum-solution-CdL3} to the vacuum EOM, one can substitute $\dot\rho$ from \eqref{Eeqd1} back into the action \eqref{CdL-action}, obtaining the following on-shell action
\begin{equation}\label{on-shell-action-CdL}
S_{\mathrm E}[\phi]  = \frac{4\pi^{\tfrac{d}{2}}}{\Gamma(\tfrac{d}{2})} \int_{0}^{\xi_{\mathrm{max}}} \mathrm{d}\xi \biggl[ \rho^{d-1}V - \frac{(d-1)(d-2)}{2\kappa_{d}} \rho^{d-3}H^2 \biggr] \, ,
\end{equation}
where we used \eqref{Sigdef}.
The bounce is then defined as
\begin{align}
B \equiv S_{\mathrm E}[\phi]-S_{\mathrm E}[\phi_{+}] = B_{\mathrm{in}}+ B_{\mathrm{wall}} + B_{\mathrm{out}} \, ,
\end{align}
where we split the total bounce $B$ into three contributions upon using that the domain of integration of the action naturally splits into three regions. These are defined in terms of the interior, the thin wall and the exterior of the bubble and we will denote by $\phi_{+}$ and $\phi_{-}$ the false and true vacuum, respectively found outside and inside the bubble of true vacuum.

On the thin wall the scalar field $\phi$ varies, while $\rho$ is considered nearly constant $\bar{\rho}$ and the potential can be approximated, for nearly degenerate vacua, by some function $V_0(\phi)$ such that {$V_0(\phi_+)=V_0(\phi_-)$, $V_0'(\phi_\pm)=0$ and $V(\phi) = V_{0}(\phi)+\mathcal{O}(\epsilon)$, with $\epsilon\equiv V_{+}-V_{-} \simeq 0$. The difference of on-shell actions then only depends on the first term in \eqref{on-shell-action-CdL}, leading to
\begin{equation}\label{Bwalld}
B_{\mathrm{wall}} \equiv S_{\mathrm E}^{\mathrm{wall}}[\phi]-S_{\mathrm E}^{\mathrm{wall}}[\phi_{+}] \simeq  \Sigma(S^{d-1})\bar{\rho}^{d-1}S_{1} \, ,
\end{equation}
where we introduced $S_1$ following \cite{Coleman:1980aw},
\begin{align}\label{S1-CdL}
S_{1} \equiv 2\!\int_{\bar{\xi}-\delta}^{\bar{\xi}+\delta}\! \mathrm{d}\xi \, [ V_{0}(\phi)-V_{0}(\phi_{+}) ]\, .
\end{align}
Here $\bar{\xi}$ identifies the center of the thin wall and $\rho$ has been approximated, on the entire wall, by its value at that point, namely $\bar{\rho}\equiv \rho(\bar{\xi})$. $\delta$ denotes an infinitesimal shift in the coordinate $\xi$, in line with the fact that the wall is assumed to be thin.

Inside and outside the bubble, the potential is constant and $\dot\phi=0$, so one can exploit \eqref{Eeqd1} to change variable in the on-shell action \eqref{on-shell-action-CdL} via
\begin{equation}
\mathrm{d}\xi = \mathrm{d}\rho \, \biggl[  H^2 - \frac{2\kappa_{d}}{(d-1)(d-2)}\rho^2 V_{c} \biggr]^{-\tfrac{1}{2}} \, ,
\end{equation}
obtaining the expressions
\begin{equation}
\begin{aligned}
S_{\mathrm E}^{\mathrm{in}}[\phi] &= -\frac{4\pi^{\tfrac{d}{2}}}{\Gamma(\tfrac{d}{2})}\frac{(d-1)(d-2)}{2\kappa_{d}} \int_{0}^{\bar{\rho}} \mathrm{d}\rho \, \rho^{d-3} \sqrt{ H^2 - \frac{2\kappa_{d}V_{c}}{(d-1)(d-2)}\rho^2 } \,,
\\
S_{\mathrm E}^{\mathrm{out}}[\phi] &= -\frac{4\pi^{\tfrac{d}{2}}}{\Gamma(\tfrac{d}{2})}\frac{(d-1)(d-2)}{2\kappa_{d}} \int_{\bar{\rho}}^{\rho_{\mathrm{max}}} \mathrm{d}\rho \, \rho^{d-3} \sqrt{ H^2 - \frac{2\kappa_{d}V_{c}}{(d-1)(d-2)}\rho^2} \,.
\end{aligned}
\end{equation}
Since outside the bubble the vacuum is the false one, by construction one has that $B_{\mathrm{out}}=0$. On the other hand, inside the bubble one can explicitly compute the integral
\begin{align}
B_{\mathrm{in}} & \equiv S_{\mathrm E}^{\mathrm{in}}[\phi_{-}] - S_{\mathrm E}^{\mathrm{in}}[\phi_{+}] 
\\
& = -\frac{4\pi^{\tfrac{d}{2}}}{\Gamma(\tfrac{d}{2})}\frac{(d-1)(d-2)}{2\kappa_{d}} \times
\notag \\
& \qquad \int_{0}^{\bar{\rho}} \mathrm{d}\rho \, \rho^{d-3} \biggl[ \sqrt{ H^2 - \frac{2\kappa_{d}V_{-}}{(d-1)(d-2)} \rho^2} - \sqrt{ H^2 - \frac{2\kappa_{d}V_{+}}{(d-1)(d-2)} \rho^2}\biggr]
\notag \\
& = -\frac{2\pi^{\tfrac{d}{2}}}{\Gamma(\tfrac{d}{2})}\frac{d-1}{\kappa_{d}}H\bar{\rho}^{d-2} \biggl[ F\biggl(-\frac{1}{2},\frac{d}{2}-1;\frac{d}{2};z_{-}\biggr)-F\biggl(-\frac{1}{2},\frac{d}{2}-1;\frac{d}{2};z_{+}\biggr) \biggr] \,\, ,
\notag
\end{align}
after having changed the variable in the last step, $\rho = \bar{\rho}\sqrt{t}$, and used the Euler-type integral representation of hypergeometric functions,
\begin{equation}
F(a,b;c;z)=\frac{\Gamma(c)}{\Gamma(b)\Gamma(c-b)}\!\int_{0}^{1}\! \mathrm{d}t \, t^{b-1}(1\!-\!t)^{c-b-1}(1\!-\!zt)^{-a} ~~ \text{for} ~~  \text{Re}(b),\,\text{Re}(c)\!>\!0 \, .
\end{equation}
In the above case we made use of the following identifications
\begin{equation}
a=-\frac{1}{2}\,,
\qquad
b=\frac{d}{2}-1\,, 
\qquad
c=\frac{d}{2}\,, 
\qquad
z_{\pm}=\frac{2\kappa_{d}V_{\pm}}{H^2(d-1)(d-2)}\bar{\rho}^2 \,.
\end{equation}
For comparison, one can explicitly compute the integral in the special case of $d=5$,
\begin{equation}
\begin{aligned}
& \int_{0}^{\bar{\rho}} \mathrm{d}\rho \, \rho^2 \sqrt{ H^2 - \frac{\kappa_{5}V_{c}}{6} \rho^2} 
\\
&=\frac{3H^4}{4\kappa_5V_{c}}\left[ 
\sqrt{\frac{6}{\kappa_5V_{c}}}\arcsin\left(\sqrt{\frac{\kappa_5V_{c}}{6}}\frac{\bar\rho}{H}\right)-\sqrt{1-\frac{\kappa_5V_{c}}{6}\frac{\bar\rho^2}{H^2}}\left(\frac{\bar\rho}{H}-\frac{\kappa_5V_{c}}{3}\frac{\bar\rho^3}{H^3}\right)
\right] \,,
\end{aligned}
\end{equation}
and in the case of $d=4$,
\begin{equation}
\int_{0}^{\bar{\rho}} \mathrm{d}\rho \, \rho \sqrt{ H^2 - \frac{\kappa_4V_{c}}{3} \rho^2} =\frac{H^3}{\kappa_4V_{c}}\left[ 
1-\left(1-\frac{\kappa_4V_{c}}{3}\frac{\bar\rho^2}{H^2}\right)^{3/2}
\right] \,.
\end{equation}
Alternatively, the hypergeometric function takes the following explicit form in $d=5$
\begin{align}
F\biggl(-\frac{1}{2},\frac{3}{2};\frac{5}{2};x^2\biggr)
&=\frac{3}{8}x^{-3}\left[\arcsin x-x(2x^2-1)\sqrt{1-x^2}\right] \,,
\end{align}
and in $d=4$
\begin{align}
F\biggl(-\frac{1}{2},1;2;x^2\biggr)
=\frac{2}{3}x^{-2}\left[1-(1-x^2)^{3/2}\right] \,.
\end{align}
The presence of the $\arcsin$ function is a general feature of the integral with odd $d$. Note also that the hypergeometric function equals unity at $x=0$ for any $d>2$.

Altogether, the total action of the bounce $B$ is given by
\begin{align}\label{BCdL}
B & = B_{\mathrm{in}}+B_{\mathrm{wall}} 
\\
& =\frac{2\pi^{\tfrac{d}{2}}}{\Gamma(\tfrac{d}{2})} \biggl( \bar{\rho}^{d-1}S_{1}-\tfrac{(d-1)H\bar{\rho}^{d-2}}{
\kappa_{d}} \biggl[ F\biggl(-\frac{1}{2},\frac{d}{2}-1;\frac{d}{2};z_{-}\biggr)-F\biggl(-\frac{1}{2},\frac{d}{2}-1;\frac{d}{2};z_{+}\biggr) \biggr] \biggr) \,,
\notag
\end{align}
and can be extremised with respect to $\bar{\rho}$ upon solving the condition
\begin{equation}\label{extremisation-CdL}
\frac{\mathrm{d}B}{\mathrm{d}\bar{\rho}}=(d-1)\bar{\rho}^{d-2}S_{1}-\tfrac{(d-1)(d-2)}{\kappa_{d}}\bar{\rho}^{d-3}  \biggl[ \sqrt{ H^2 - \tfrac{2\kappa_{d}\bar{\rho}^2 V_{-}}{(d-1)(d-2)} } - \sqrt{ H^2 - \tfrac{2\kappa_{d}\bar{\rho}^2 V_{+}}{(d-1)(d-2)} }\biggr] = 0 \, .
\end{equation}

We now consider the two special cases studied in \cite{Coleman:1980aw}, namely $V_{+}=\epsilon, \, V_{-}=0$ and $V_{+}=0, \, V_{-}=-\epsilon$, respectively corresponding to transitions from dS to Minkowski and from Minkowski to AdS. To better understand and visualise the final expressions for the bounce action $B$, it will turn out useful to express it, both here and in section \ref{sec:examples}, in terms of the following dimensionless quantity
\begin{equation}\label{dimensionless-kappa}
\hat{\kappa}_{d} \equiv \kappa_{d} \frac{S_{1}^2}{\epsilon} \,.
\end{equation}
In all considered cases, the effect of introducing this quantity is to recast complicated functions $B=B(H,\kappa_{d},S_{1},\epsilon)$ in the form
\begin{equation}
B(H,\kappa_{d},S_{1},\epsilon) =  F(H,S_{1},\epsilon) \mathcal{B}(\hat{\kappa}_{d}) \,,
\end{equation}
with $F(H,S_{1},\epsilon)$ a simple dimensionless ratio of $H,S_{1},\epsilon$ and $\mathcal{B}(\hat{\kappa}_{d})$ a dimensionless function of $\hat{\kappa}_{d}$ only. Both quantities change across dimensions and symmetry of the bounce, while only $\mathcal{B}(\hat{\kappa}_{d})$ further depends on the type of transition.
This rewriting allows for an easier evaluation of the gravity decoupling limit $\kappa_{d}\rightarrow 0$, which corresponds to $\hat{\kappa}_{d}\rightarrow 0$, and of gravitational corrections. Furthermore, $\hat{\kappa}_{d}$ is the only dimensionless combination of $\kappa_{d},S_{1},\epsilon$ which allows for a series expansion of $B$ in powers of $\kappa_{d}$ that is also analytic in $S_{1}$. Finally, the function $F$ can be chosen so as to normalise $\mathcal{B}(\hat{\kappa}_{d})$ to 1 in the $\hat{\kappa}_{d}\rightarrow 0$ limit.
\begin{itemize}
\item \textbf{dS $\rightarrow$ Minkowski.} \qquad Setting $V(\phi_{+})=\epsilon$ and $V(\phi_{-})=0$ simplifies \eqref{extremisation-CdL} as
\begin{equation}
\kappa_{d}S_{1}\bar{\rho}-(d-2)H+(d-2)\sqrt{H^2 -\frac{2\kappa_{d}\bar{\rho}^2\epsilon}{(d-1)(d-2)} } = 0 \,,
\end{equation}
which is solved by
\begin{equation}\label{CdL-V+=epsiolon-V-=0-solution}
\bar{\rho} = \frac{2(d-1)(d-2)S_{1}H}{2(d-2)\epsilon + (d-1)\kappa_{d}S_{1}^2} \, .
\end{equation}
Substituting back into $B$ and introducing $\hat{\kappa}_{d}$ as in \eqref{dimensionless-kappa} one obtains the expressions
\begin{equation}\label{Bnoncompact}
B_{\mathrm{dS}}^{4\mathrm d} = \frac{27\pi^2 S_{1}^4}{2\epsilon^3} \mathcal{B}_{\mathrm{dS}}^{4\mathrm d}(\hat{\kappa}_{4}) \qquad \text{and} \qquad 
B_{\mathrm{dS}}^{5\mathrm d}= \frac{2048 \pi^2 S_{1}^5}{15 \epsilon^4}\mathcal{B}_{\mathrm{dS}}^{5\mathrm d}(\hat{\kappa}_{5}) 
\end{equation}
with the dimensionless functions exhibiting the following behaviour for $\hat{\kappa}_{d}\rightarrow 0$:
\begin{equation}
\begin{aligned}
\mathcal{B}_{\mathrm{dS}}^{4d}(\hat{\kappa}_{4})&=\frac{16}{(4+3\hat{\kappa}_{4})^2}\simeq 1-\frac{3}{2}\hat{\kappa}_{4}+\frac{27}{16}\hat{\kappa}_{4}^2 +\mathcal{O}(\hat{\kappa}_{4}^3) \,,
\\
\mathcal{B}_{\mathrm{dS}}^{5d}(\hat{\kappa}_{5})&\simeq 1-\frac{40}{21}\hat{\kappa}_{5}+\frac{200}{81}\hat{\kappa}_{5}^2+\mathcal{O}(\hat{\kappa}_{5}^3) \,.
\end{aligned}
\end{equation}
Note that the bounce action is independent of $H$ in any dimension and in the absence of gravity is proportional to the dimensionless ratio $S_1^d/\epsilon^{d-1}$.
\item \textbf{Minkowski $\rightarrow$ AdS.} \qquad Setting $V(\phi_{+})=0$ and $V(\phi_{-})=-\epsilon$ simplifies \eqref{extremisation-CdL} as
\begin{equation}
\kappa_{d}S_{1}\bar{\rho}+(d-2)H-(d-2)\sqrt{H^2 +\frac{2\kappa_{d}\bar{\rho}^2\epsilon}{(d-1)(d-2)} } = 0 \, ,
\end{equation}
which is solved by
\begin{equation}\label{CdL-V+=0-V-=-epsilon-solution}
\bar{\rho} = \frac{2(d-1)(d-2)S_{1}H}{2(d-2)\epsilon - (d-1)\kappa_{d}S_{1}^2} \, .
\end{equation}
Proceeding as above one finds the following expressions
\begin{equation}
B_{\mathrm{AdS}}^{4d} = \frac{27\pi^2 S_{1}^4}{2\epsilon^3} \mathcal{B}_{\mathrm{AdS}}^{4d}(\hat{\kappa}_{4}) \qquad \text{with} \qquad 
B_{\mathrm{AdS}}^{5d}= \frac{2048 \pi^2 S_{1}^5}{15 \epsilon^4}\mathcal{B}_{\mathrm{AdS}}^{5d}(\hat{\kappa}_{5})
\end{equation}
with the dimensionless functions exhibiting the following behaviour for $\hat{\kappa}_{d}\rightarrow 0$
\begin{equation}
\begin{aligned}
\mathcal{B}_{\mathrm{AdS}}^{4d}(\hat{\kappa}_{4})&=\frac{16}{(4-3\hat{\kappa}_{4})^2}\simeq 1+\frac{3}{2}\hat{\kappa}_{4}+\frac{27}{16}\hat{\kappa}_{4}^2 +\mathcal{O}(\hat{\kappa}_{4}^3) \,,
\\
\mathcal{B}_{\mathrm{AdS}}^{5d}(\hat{\kappa}_{5})&\simeq 1+\frac{40}{21}\hat{\kappa}_{5}+\frac{200}{81}\hat{\kappa}_{5}^2+\mathcal{O}(\hat{\kappa}_{5}^3) \,.
\end{aligned}
\end{equation}
Actually, it is easy to see from \eqref{BCdL} and \eqref{extremisation-CdL} that $B_{\mathrm AdS}(\kappa_d,\epsilon)=B_{\mathrm dS}(-\kappa_d,-\epsilon)$ and thus $\mathcal{B}_{\mathrm{AdS}}(\hat{\kappa}_{d})=\mathcal{B}_{\mathrm{dS}}(-\hat{\kappa}_{d})$ in all dimensions.
\end{itemize}

\subsection{The approach of Israel junction conditions}
\label{subsec:israel-CdL}
In this subsection we present a different derivation of the above results, based on the so-called Israel junction conditions. This approach takes a more geometrical perspective on the problem and relies on the emergence of continuity conditions to be imposed on the spacetime variables, rather than on extremisation procedures. The main advantage of this picture is that it provides a clearer way of proceeding in more complicated settings, such as the one considered in the next section, which is the main purpose of this work.

Let us assume that $V$ has two minima at $\phi_\pm$ with potential values $V_\pm$ satisfying $V_+-V_-=\ep$.
Suppose that $\phi$ is at the minimum on each side of the wall: $\phi=\phi_+$ outside and $\phi=\phi_-$ inside, while on the wall $\phi$ depends only on $\xi$.

We consider the Euclidean spacetime to be separated by a thin wall, which is a hypersurface $\xi=\bar\xi$, associated to the world-volume of a $(d-2)$-brane with tension $T$. 
We suppose that the metrics inside and outside of the wall have the form \eqref{gtilde-metric-CdL}.
The induced metrics on the two sides of the thin wall are $\rho(\bar\xi\pm0)^2g_{ij}\mathrm{d}y^i \mathrm{d}y^j$. We first impose that they should coincide:
\begin{align}
\bar\rho\equiv\rho(\bar\xi+0)=\rho(\bar\xi-0) \,.
\end{align}
Next, let us derive a junction condition for the derivative of $\rho$ at the wall. 
Recall that the Ricci tensor for $g$ can be written as
\begin{align}
R[\tilde g]_{ij}=R[g]_{ij}-(d-2)\dot\rho^2g_{ij}-\frac{1}{2}\frac{\mathrm{d}^2(\rho^2)}{\mathrm{d}\xi^2}g_{ij} \,.
\end{align}
Integrating this over a tiny region $[\bar\xi-\de,\bar\xi+\de]$,\footnote{$\de$ is taken such that it is sufficiently small while the region is slightly thicker than the thin wall.} we obtain
\begin{align}
\int_{\bar\xi-\de}^{\bar\xi+\de}\mathrm{d}\xi\,R[\tilde g]_{ij}=-\frac{1}{2}g_{ij}\frac{\mathrm{d}(\rho^2)}{\mathrm{d}\xi}\Big|_{\bar\xi+\de}+\frac{1}{2}g_{ij}\frac{\mathrm{d}(\rho^2)}{\mathrm{d}\xi}\Big|_{\bar\xi-\de} \,.
\end{align}
Replacing the components of the Ricci tensor with $R_{\mu\nu}=\kappa_d(2g_{\mu\nu}V/(d-2)+\pd_\mu\phi\pd_\nu\phi)$ that comes from the Einstein equation, and noting also that $\phi$ depends only on $\xi$, we obtain
\begin{align}
\kappa_d\int_{\bar\xi-\de}^{\bar\xi+\de}\mathrm{d}\xi\,\left[\pd_i\phi\pd_j\phi+\frac{2}{d-2}V\rho^2g_{ij}\right]=-\frac{1}{2}g_{ij}\frac{\mathrm{d}(\rho^2)}{\mathrm{d}\xi}\Big|_{\bar\xi+\de}+\frac{1}{2}g_{ij}\frac{\mathrm{d}(\rho^2)}{\mathrm{d}\xi}\Big|_{\bar\xi-\de} \,.
\end{align}
On our on-shell configurations, it becomes
\begin{align}\label{israel-d}
\frac{1}{d-2}\kappa_d\bar\rho^2g_{ij}\int_{\bar\xi-0}^{\bar\xi+0}\mathrm{d}\xi\,2V=-g_{ij}\bar\rho\dot\rho_++g_{ij}\bar\rho\dot\rho_- \,,
\end{align}
where we introduced $\dot\rho_\pm \equiv \dot\rho(\bar\xi\pm0)$.
The integral of $2V$ in \eqref{israel-d} can be approximated as $S_1$ in \eqref{S1-CdL} since the difference is of higher order in $\ep$,
\begin{align}\label{int2V=S1}
\int_{\bar\xi-0}^{\bar\xi+0}\mathrm{d}\xi\,2V \simeq S_1 \,.
\end{align}
The junction condition then becomes
\begin{align}\label{jump-d}
\frac{1}{d-2}\kappa_dS_1\bar\rho^2=-\bar\rho\dot\rho_++\bar\rho\dot\rho_- \,,
\end{align}
and $S_1/2$ can be identified with the brane tension $T$.

Let us apply these general junction conditions to the two concrete cases considered in the previous subsection by using the general solutions \eqref{Vacuum-solution-CdL1}--\eqref{Vacuum-solution-CdL3}.
\begin{itemize}
\item \textbf{dS $\rightarrow$ Minkowski.} \qquad In this case $V_+=\ep$ and $V_-=0$ with $\ep>0$, for which 
\begin{alignat}{2}
\rho(\xi)&=H\ell\sin\left(\frac{\xi-\xi_0}{\ell}\right) &\qquad \mbox{for} ~~ &\xi>\bar\xi \,, \\
\rho(\xi)&=H\xi &\qquad \mbox{for} ~~ &\xi<\bar\xi \,,
\end{alignat}
where $\xi_{0}$ is an integration constant and we introduced 
\begin{align}\label{ell-def}
\ell=\sqrt{\frac{(d-1)(d-2)}{2\kappa_d\ep}} \,.
\end{align}
We first have the continuity of $\rho$ at the wall:
\begin{align}
\bar\rho=H\bar\xi=H\ell\sin\left(\frac{\bar\xi-\xi_0}{\ell}\right) \,.
\end{align}
The jumping condition \eqref{jump-d}, combined with this, becomes
\begin{align}
-\cos\left(\frac{\bar\xi-\xi_0}{\ell}\right)+1=\frac{1}{d-2}\kappa_dS_1\frac{\bar\rho}{H} \,.
\end{align}
It follows that
\begin{align}
1-\sqrt{1-\frac{\bar\rho^2}{(H\ell)^2}}=\frac{1}{d-2}\kappa_dS_1\frac{\bar\rho}{H} \,, 
\end{align}
and solving for $\bar\rho$ upon substituting \eqref{ell-def} for $\ell$, we recover \eqref{CdL-V+=epsiolon-V-=0-solution},
\begin{align}
\bar\rho=\frac{2(d-1)(d-2)S_1H}{2(d-2)\ep+(d-1)\kappa_dS_1^2} \,.
\end{align}
\item \textbf{Minkowski $\rightarrow$ AdS.} \qquad In this case $V_+=0$ and $V_-=-\ep$ with $\ep>0$, for which
\begin{alignat}{2}
\rho(\xi)&=H(\xi-\xi_0) &\qquad \mbox{for} ~~ &\xi>\bar\xi \,, \\
\rho(\xi)&=H\ell\sinh\left(\frac{\xi}{\ell}\right) &\qquad \mbox{for} ~~ &\xi<\bar\xi \,,
\end{alignat}
where $\ell$ is \eqref{ell-def}.
We first have the continuity of $\rho$ at the wall:
\begin{align}
\bar\rho=H(\bar\xi-\xi_0)=H\ell\sinh\left(\frac{\bar\xi}{\ell}\right) \,.
\end{align}
The jumping condition \eqref{jump-d}, combined with this, becomes
\begin{align}
-1+\cosh\left(\frac{\bar\xi}{\ell}\right)=\frac{1}{d-2}\kappa_dS_1\frac{\bar\rho}{H} \,.
\end{align}
It follows that
\begin{align}
\sqrt{1+\frac{\bar\rho^2}{(H\ell)^2}}-1=\frac{1}{d-2}\kappa_dS_1\frac{\bar\rho}{H} \,,
\end{align}
and solving for $\bar\rho$ upon substituting \eqref{ell-def} for $\ell$, we recover \eqref{CdL-V+=0-V-=-epsilon-solution},
\begin{align}
\bar\rho=\frac{2(d-1)(d-2)S_1H}{2(d-2)\ep-(d-1)\kappa_dS_1^2} \,.
\end{align}
\end{itemize}

\section{Vacuum Decay in Theories with $O(d-1)\times U(1)$ Symmetry}\label{sec:case-of-compact-dimensions}
In this section we turn our attention to the slightly more complicated setting of a background enjoying $O(d-1)\times U(1)$ symmetry. The physical motivation for considering such a scenario comes from the need for a description of vacuum transitions in the case of geometries with compact dimensions, which arise, for example, in stringy or Kaluza-Klein-inspired cosmological or particle physics models.
The starting point is once again the Euclidean action \eqref{Euclidean-action}, this time equipped with the following background metric
\begin{equation}\label{metric-compact-directions-generic}
\begin{aligned}
\tilde{g} &=\mathrm{d}\xi^2+\rho(\xi)^2 [ \al^2 \mathrm{d}\Omega_{d-2}^2+\sig^2(\xi) \mathrm{d}\chi^2 ] \,,
\end{aligned}
\end{equation}
where $\mathrm{d}\Omega_{d-2}^2$ is the surface element of a $(d-2)$-dimensional unit sphere, $\mathrm{d}\chi^2$ the line element along the compact dimension and $\alpha$ is a constant carrying the unit of length and related to the radius of the sphere. Having in mind the special case of a five-dimensional spacetime with one compact dimension, we shall restrict our analysis to the case of $d=5$. The case of arbitrary $d$ should proceed along the same lines. 

Specialising to $d=5$ the explicit metric can be written as
\begin{align}
\tilde{g}=\mathrm{d}\xi^2+\al^2\rho(\xi)^2(\mathrm{d}\te_1^2+\sin^2\te_1\mathrm{d}\te_2^2+\sin^2\te_1\sin^2\te_2\mathrm{d}\varphi^2)+\rho(\xi)^2\sig(\xi)^2\mathrm{d}\chi^2 \, ,
\end{align}
containing spheres $S^{3}$ and $S^{1}$.
We adopt the following mass dimensions
\begin{equation}\label{mass-dimensions}
[\xi] \!=\! -1, \quad [\rho]\!=\!0, \quad [\sigma] \!=\! -1, \quad [\chi]\!=\!0, \quad [\theta_1]\!=\!0, \quad [\theta_2]\!=\!0, \quad [\varphi]\!=\!0, \quad [\al]\!=\!-1 \,.
\end{equation}
Here $\alpha\rho$ is the radius of the 3-sphere and $\rho\sigma$ is the radius of the circle, both depending on the radial coordinate $\xi$.

The Einstein equations of motion with matter read
\begin{align}
&\frac{6\dot\rho^2}{\rho^2}+\frac{3\dot\rho\dot\sig}{\rho\sig}-\frac{3}{\alpha^2\rho^2}=\kappa_d\bigg(\frac{\dot\phi^2}{2}-V\bigg)\,, \label{EeqS1-1} \\
&\frac{3\ddot\rho}{\rho}+\frac{\ddot\sig}{\sig}+\frac{4\dot\rho\dot\sig}{\rho\sig}+\frac{3\dot\rho^2}{\rho^2}-\frac{1}{\alpha^2\rho^2}=-\kappa_d\bigg(\frac{\dot\phi^2}{2}+V\bigg)\,, \label{EeqS1-2} \\
&\frac{3\ddot\rho}{\rho}+\frac{3\dot\rho^2}{\rho^2}-\frac{3}{\alpha^2\rho^2}=-\kappa_d\bigg(\frac{\dot\phi^2}{2}+V\bigg) \,, \label{EeqS1-3} 
\end{align}
where, once again, the single and double dot respectively denote $\mathrm{d}/\mathrm{d}\xi$ and $\mathrm{d}^2/\mathrm{d}\xi^2$.
The first equation \eqref{EeqS1-1} is the Friedmann equation in the present setup.
The equation of motion for $\phi$ reads
\begin{align}\label{EoMphi-S1}
\frac{\mathrm{d}}{\mathrm{d}\xi}(\rho^4\sig\dot\phi)=\rho^4\sig \frac{\mathrm{d} V(\phi)}{\mathrm{d}\phi}  \,,
\end{align}
and following steps similar to Section~\ref{sec:theories-with-O(d)-symmetry} the action \eqref{Euclidean-action} takes the explicit form
\begin{equation}\label{Action-physical-time-5d}
S_{\mathrm E}[\phi]\!=\! 2\pi^2\al^3 \! \int_{0}^{2\pi} \mathrm{d}\chi  \int_{0}^{\xi_{\mathrm{max}}} \mathrm{d}\xi  \left[-\frac{3}{\kappa_{5}}(\al^{-2}\rho^2\sig\!+\!2\rho^2\dot\rho^2\sig\!+\!\rho^3\dot\rho\dot\sig)\!+\!\rho^4\sig\bigg(\frac{\dot\phi^2}{2}+V\bigg)\right] \,.
\end{equation}
From now on we shall set $\alpha^2 = (2H^2)^{-1}$, with $H$ carrying mass dimension one. With this choice the curvature of $S^{4}$ with radius $H^{-1}$ coincides with the curvature of $S^{3}$ with radius $(\sqrt{2}H)^{-1}$, allowing for a convenient comparison with the non-compact case.

\subsection{Instanton geometry}
As in the CdL case, we will solve the equations of motion \eqref{EeqS1-1}--\eqref{EeqS1-3} around a local vacuum for a constant configuration of $\phi$ and a constant potential $V_c$. Multiplying \eqref{EeqS1-3} by $\rho^3\dot\rho$, we find a total derivative,
\begin{align}
\frac{\mathrm{d}}{\mathrm{d}\xi}\left(\frac{3}{2}\rho^2\dot\rho^2-3H^2\rho^2+\frac{\kappa_5V_c}{4}\rho^4\right)=0 \,,
\end{align}
which, with an integration constant $\bt$ carrying dimensions $[\beta]=2$, yields
\begin{align}\label{rhoeqbetaS1}
\frac{3}{2}\dot\rho^2-3H^2+\frac{\kappa_5V_c}{4}\rho^2=\frac{\bt}{\rho^2} \,.
\end{align}
On the other hand, the difference \eqref{EeqS1-1}$-$\eqref{EeqS1-3} gives another total derivative,
\begin{align}
\frac{3\dot\rho^2}{\rho^2\sig}\frac{\mathrm{d}}{\mathrm{d}\xi}\bigg(\frac{\rho\sig}{\dot\rho}\bigg)=0 \,,
\end{align}
which is solved with an integration constant $\gamma$ carrying dimensions $[\gamma]=-2$, by
\begin{align}\label{sigeq}
\rho\sig=\gamma\dot\rho \,.
\end{align}
Therefore, once we solve \eqref{rhoeqbetaS1}, we can obtain $\sig$ by \eqref{sigeq}. Note that an overall constant in $\sig$ cannot be fixed because the Einstein equations are invariant under the rescaling of $\sig$.

At this point we stress that only the function $\rho$ is left undetermined in the above system of equations, and before proceeding
we further notice that since the solution to the EOM depends on the choice of constant potential $V_{c}$, it makes sense to make distinction between the two vacua $V_{\pm}$ by rewriting \eqref{sigeq} as two independent equations
\begin{equation}
\rho_{\pm}\sigma_{\pm} = \gamma_{\pm}\dot\rho_{\pm} \, .
\end{equation}
Before fixing $V_c$, one can solve the integrated equation \eqref{rhoeqbetaS1} by first separating variables
\begin{equation}
\frac{(\mathrm{d}\rho)^2}{2H^2-\frac{\kappa_{5}V_{c}\rho^2}{6}+\frac{2\beta}{3\rho^2}}=(\mathrm{d}\xi)^2
\end{equation}
and further rearranging as
\begin{equation}
\frac{\mathrm{d}x}{\sqrt{ax^2+bx +c}} = \mathrm{d}\xi
\qquad 
\text{with} 
\qquad
x\equiv \rho^2 \quad a \equiv - \frac{2\kappa_{5}V_{c}}{3}
\quad b\equiv 8H^2
\quad c\equiv \frac{8\beta}{3} \, .
\end{equation}
In the latter form, the equation can be integrated explicitly and the resulting expression depends on the signs of $a,b,c$ and $\Delta \equiv \frac{1}{4}b^2-ac$, which must make the polynomial positive for the square root to be real. The following cases will be relevant for our discussion:
\begin{align}\label{vacuum-solution-4-cases}
& \textbf{dS.} \quad \quad \,\,  a<0 \quad \text{and} \quad \Delta > 0 \quad  \Rightarrow \quad \frac{-1}{\sqrt{-a}}\arcsin{\biggl( \frac{2ax + b}{\sqrt{4\Delta}} \biggr)} = \xi + \xi_{0}
\notag \\
& \textbf{Mink.} \quad  \qquad \quad a=0 \qquad \quad \quad \,\,  \Rightarrow \quad  \frac{2}{b}\sqrt{bx +c} = \xi + \xi_{0}
\notag \\
\\
& \textbf{AdS1.} \quad a>0 \quad \text{and} \quad \Delta > 0 \quad \Rightarrow \quad \frac{1}{\sqrt{a}}\log{\biggl( 2\sqrt{a(ax^2+bx+c)}+2ax+b \biggr)} = \xi +\xi_{0}
\notag \\ 
& \textbf{AdS2.} \quad a>0 \quad \text{and} \quad \Delta < 0 \quad \Rightarrow \quad \frac{1}{\sqrt{a}} \arcsinh{\biggl( \frac{2ax+b}{\sqrt{-4\Delta}} \biggr)} = \xi +\xi_{0}
\notag \\
& \textbf{AdS3.}  \quad a>0 \quad \text{and} \quad \Delta = 0 \quad \Rightarrow \quad 
\frac{1}{\sqrt{a}}\log{\Bigl( x+\frac{b}{2a} \Bigr)} = \xi+\xi_{0}
\notag
\end{align}
Varying the signs of $V_{c}$ and $\beta$, the latter being at this level an undetermined integration constant, we will fall in each of the above cases. We define the roots of the polynomial as
\begin{equation}
x_{\pm} \equiv -\frac{b}{2a}\mp \frac{1}{a}\sqrt{\Delta}= \frac{3\bigl[4H^2 \pm \sqrt{\Delta}\bigr]}{2\kappa_{5}V_{c}}\qquad \text{with} \qquad \Delta=16(H^4+\frac{1}{9}\kappa_{5}V_{c}\beta)
\end{equation}
and recall that we also have the requirement $x\equiv\rho^2 \geq 0$. This leads to the following solutions:
\begin{itemize}
\item \textbf{dS.} \quad For $V_{c} > 0$ one has $a < 0$, and the polynomial can only be positive for
\begin{equation}
\Delta  > 0 \qquad \Rightarrow \qquad  -\frac{9H^4}{\kappa_{5}|V_{c}|}  < \beta \, ,
\end{equation}
within the interval $x_{-} \le x \le x_{+}$. Given also the requirement $x\equiv \rho^2 \geq 0$ one further needs to make sure that $x_{-} \geq 0$. This condition is satisfied for 
\begin{equation}
4H^2 - \sqrt{16H^4+\frac{16}{9}\kappa_{5}|V_{c}|\beta} \,  \geq 0 \qquad \Rightarrow \qquad \beta \le 0 \,.
\end{equation}
Altogether, a positive potential implies the EOM can be solved provided that
\begin{equation}
-\frac{9H^4}{\kappa_{5}|V_{c}|} < \beta \le 0 \, .
\end{equation}
The explicit form of the solution then reads
\begin{equation}\label{Vacuum-solution-Vc>0}
\begin{aligned}
\rho(\xi) &= \sqrt{\frac{6H^2}{\kappa_{5}|V_{c}|}+\frac{6\sqrt{H^4+\tfrac{1}{9}\kappa_{5}|V_{c}|\beta}}{\kappa_{5}|V_{c}|}\,\, \sin\biggl[ \sqrt{\frac{2\kappa_{5}|V_{c}|}{3}}\xi-\xi_{0} \biggr]} \,,
\\
\\
\sigma(\xi) & = \gamma \,\,  \frac{\sqrt{\frac{\kappa_{5}|V_{c}|}{6}}\sqrt{H^4+\tfrac{1}{9}\kappa_{5}|V_{c}|\beta} \,\, \cos\biggl[ \sqrt{\frac{2\kappa_{5}|V_{c}|}{3}}\xi-\xi_{0} \biggr]}{H^2+\sqrt{H^4+\tfrac{1}{9}\kappa_{5}|V_{c}|\beta}\,\, \sin\biggl[ \sqrt{\frac{2\kappa_{5}|V_{c}|}{3}}\xi-\xi_{0} \biggr]} \,.
\end{aligned}
\end{equation}
\item \textbf{Mink.} \quad For $V_{c}=0$ one has $a=0$. The polynomial is positive for $x \geq -\frac{c}{b} = -\frac{\beta}{3H^2}$ and given the requirement $x\equiv \rho^2 \geq 0$ one finds the condition $\beta \le 0$. In this regime, the solution reads
\begin{equation}\label{Vacuum-solution-V=0}
\begin{aligned}
\rho(\xi)&=\sqrt{2H^2(\xi-\xi_{0})^2-\frac{\beta}{3H^2}} \,,
\\
\sigma(\xi)&=-\gamma \,\, \frac{6H^4(\xi-\xi_{0})}{\beta - 6H^4(\xi-\xi_{0})^2} \,.
\end{aligned}
\end{equation}
\item \textbf{AdS1.} \quad  For $V_{c}<0$ one has $a>0$, and from \eqref{vacuum-solution-4-cases} the first case to consider is
\begin{equation}\label{condition_Delta_case3}
\Delta  > 0 \qquad \Rightarrow \qquad \beta  < \frac{9H^4}{\kappa_{5}|V_{c}|} \, .
\end{equation}
In this case the roots are such that $x_{+}<x_{-}$ and the polynomial is positive for $x\le x_{+}$ and $x \geq x_{-}$. The root $x_{+}$ is however always negative and the requirement $x\equiv \rho^2 \geq 0$ forces its exclusion. Making sure that $x_{-}\geq 0$ then requires
\begin{equation}
4H^2 - \sqrt{16H^4-\frac{16}{9}\kappa_{5}|V_{c}|\beta} \,   \le 0 \qquad \Rightarrow \qquad \beta \le 0 \,.
\end{equation}
The latter condition is more restrictive than \eqref{condition_Delta_case3}, hence one should consider the domain $\beta \le 0$.
After dividing the argument of the $\log$ by $H^2$ to make it dimensionless, the explicit form of the solution reads
\begin{equation}\label{Case3-general-solution}
\begin{aligned}
\rho(\xi)&=\frac{1}{2\sqrt{6\kappa_{5}|V_{c}|}}e^{-\sqrt{\frac{\kappa_{5}|V_{c}|}{6}}\xi-\xi_{0}} \sqrt{9H^4\biggl( e^{2\sqrt{\frac{\kappa_{5}|V_{c}|}{6}}\xi+2\xi_{0}} - 8  \biggr)^2 - 64\kappa_{5}|V_{c}|\beta} \,,
\\
\\
\sigma(\xi)&= \gamma \sqrt{\frac{\kappa_{5}|V_{c}|}{6}} \, \, \frac{9H^4 e^{4\sqrt{\frac{\kappa_{5}|V_{c}|}{6}}\xi+4\xi_{0}}  + 64\kappa_{5}|V_{c}|\beta-576 H^4}{9H^4\biggl( e^{2\sqrt{\frac{\kappa_{5}|V_{c}|}{6}}\xi+2\xi_{0}} - 8  \biggr)^2 - 64\kappa_{5}|V_{c}|\beta} \,.
\end{aligned}
\end{equation}
\item \textbf{AdS2.} \quad Still for $V_{c}<0$, which implies $a>0$, from \eqref{vacuum-solution-4-cases} the second possibility is
\begin{equation}
\Delta  < 0 \qquad \Rightarrow \qquad \beta  > \frac{9H^4}{\kappa_{5}|V_{c}|} \, .
\end{equation}
In this case the polynomial is always positive and no restriction on the integration constant $\beta$ is imposed. The explicit form of the solution reads
\begin{equation}\label{Case4-general-solution}
\begin{aligned}
\rho(\xi)&=\sqrt{\frac{2}{\kappa_{5}|V_{c}|}}\;\biggl[ \sqrt{-9H^4+\kappa_{5}|V_{c}|\beta} \, \sinh{\biggl( \sqrt{\frac{2\kappa_{5}|V_{c}|}{3}}(\xi+\xi_{0}) \biggr)} -3H^2 \biggr]^{1/2} \,,
\\
\\
\sigma(\xi)&=\gamma \sqrt{\frac{\kappa_{5}|V_{c}|}{6}} \,\, \frac{\sqrt{-9H^4+\kappa_{5}|V_{c}|\beta} \, \cosh{\biggl( \sqrt{\frac{2\kappa_{5}|V_{c}|}{3}}(\xi+\xi_{0}) \biggr)}}{\sqrt{-9H^4+\kappa_{5}|V_{c}|\beta} \, \sinh{\biggl( \sqrt{\frac{2\kappa_{5}|V_{c}|}{3}}(\xi+\xi_{0}) \biggr)} -3H^2} \,.
\end{aligned}
\end{equation}
\item \textbf{AdS3.} \quad $\Delta=0$ can be obtained by fixing $\beta=-\frac{9H^2}{\kappa_{5}V_{c}}$ and for the polynomial to be positive $a$ needs to be positive as well. This means that $V_{c}<0$. The solution reads
\begin{equation}\label{sol.AdS3}
\begin{aligned}
\rho(\xi)&=\sqrt{e^{\sqrt{\frac{2\kappa_{5}|V_{c}|}{3}}\xi+\xi_{0}}-\frac{6H^2}{\kappa_{5}|V_{c}|}} \,,
\\
\sigma(\xi)&=\frac{\gamma\sqrt{\frac{2\kappa_{5}|V_{c}|}{3}}e^{\sqrt{\frac{2\kappa_{5}|V_{c}|}{3}}\xi+\xi_{0}}}{2\left({e^{\sqrt{\frac{2\kappa_{5}|V_{c}|}{3}}\xi+\xi_{0}}-\frac{6H^2}{\kappa_{5}|V_{c}|}}\right)} \,.
\end{aligned}
\end{equation}
\end{itemize}

\subsection{Bounce and complications of the CdL approach} 
At this point, one may consider proceeding with the extremisation approach of Coleman and de Luccia. In analogy with their argument, presented in section \ref{sec:theories-with-O(d)-symmetry}, the action \eqref{Action-physical-time-5d} can be simplified by exploiting the EOM \eqref{EeqS1-1},
\begin{equation}
S_{\mathrm E}[\phi]=\frac{4\pi^2\alpha^3}{\kappa_{5}}\int_{0}^{2\pi} \!\mathrm{d}\chi  \int_{0}^{\xi_{\mathrm{max}}} \!\mathrm{d}\xi \,\, \rho^4\sigma \biggl( \kappa_{5}V-\frac{6H^2}{\rho^2} \biggr) \,,
\end{equation}
and the bounce is once again defined to be
\begin{equation}
B\equiv S_{\mathrm{E}}[\phi]-S_{\mathrm{E}}[\phi_{+}] = B_{\mathrm{in}} + B_{\mathrm{wall}} + B_{\mathrm{out}} \, .
\end{equation}
On the thin-wall one now proceeds by approximating not only $\rho$ but also $\sigma$, as $\rho\simeq \bar{\rho}$ and $\sigma\simeq \bar{\sigma}$, with $\bar{\rho}\equiv \rho(\bar{\xi})$ and $\bar{\sigma}\equiv \sigma(\bar{\xi})$. The potential is also approximated, as in CdL, by some function $V_0(\phi)$ such that $V(\phi)=V_{0}(\phi)+\mathcal{O}(\epsilon)$, for $\epsilon\equiv V_{+}-V_{-}\simeq 0$. This leads to
\begin{equation}\label{BwallS1_S1}
B_{\mathrm{wall}} \simeq  2\pi^2\alpha^3 \bar{\rho}^4\bar{\sigma}S_{1}\int_{0}^{2\pi} \mathrm{d}\chi \, ,
\end{equation}
where $S_1$ is the one defined in \eqref{S1-CdL}.
Outside the bubble $B_{\mathrm{out}}=0$ by construction, while inside one finds the expression
\begin{equation}
B_{\mathrm{in}}=\frac{4\pi^2\alpha^3}{\kappa_{5}}\int_{0}^{2\pi} \mathrm{d}\chi \int_{0}^{\bar{\xi}} \mathrm{d}\xi \biggl( \rho_{-}^4\sigma_{-}\biggl[ \kappa_{5}V_{-}-\frac{6H^2}{\rho_{-}^2}  \biggr] - \rho_{+}^4\sigma_{+}\biggl[ \kappa_{5}V_{+}-\frac{6H^2}{\rho_{+}^2}  \biggr] \biggr) \,,
\end{equation}
with $\rho_{\pm}$ and $\sigma_{\pm}$ respectively solving the EOM for $V_{\pm}$ and related by $\rho_{\pm}\sigma_{\pm} = \gamma_{\pm}\dot\rho_{\pm}$, with $\gamma_{\pm}$ arbitrary integration constants. Exploiting the latter relation and changing variables $\mathrm{d}\xi\, \dot\rho_{\pm} = \mathrm{d}\rho_{\pm}$ one obtains
\begin{align}
B_{\mathrm{in}}&=\frac{4\pi^2\alpha^3}{\kappa_{5}}\int_{0}^{2\pi} \mathrm{d}\chi  \biggl( \int_{0}^{\bar{\rho}} \mathrm{d}\rho_{-} \gamma_{-}\rho_{-}^3 \biggl[ \kappa_{5}V_{-}-\frac{6H^2}{\rho_{-}^2}  \biggr] - \int_{0}^{\bar{\rho}} \mathrm{d}\rho_{+} \gamma_{+}\rho_{+}^3\biggl[ \kappa_{5}V_{+}-\frac{6H^2}{\rho_{+}^2}  \biggr] \biggr)
\notag \\
& =\frac{4\pi^2\alpha^3}{\kappa_{5}}\int_{0}^{2\pi} \mathrm{d}\chi \,\, \biggl( \gamma_{-}\biggl[ \frac{\kappa_{5}V_{-}}{4}\rho_{-}^{4}-3H^2\rho_{-}^2\biggr]_{0}^{\bar{\rho}}-\gamma_{+}\biggl[ \frac{\kappa_{5}V_{+}}{4}\rho_{+}^{4}-3H^2\rho_{+}^2\biggr]_{0}^{\bar{\rho}} \biggr) \, ,
\end{align}
which leads to
\begin{align}
B_{\mathrm{in}}=\frac{2\pi^2\alpha^3}{\kappa_{5}} \int_{0}^{2\pi} \mathrm{d}\chi & \biggl[ \,\,  6( \gamma_{+}-\gamma_{-}) H^2\bar{\rho}^2 - \frac{\kappa_{5}}{2}(V_{+}\gamma_{+}-V_{-}\gamma_{-})\bar{\rho}^4 + 
\\
& - 6 \Bigl( \gamma_{+}\rho_{+}^2(0)-\gamma_{-}\rho_{-}^2(0)\Bigr)H^2+\frac{\kappa_{5}}{2}\Bigl( V_{+}\gamma_{+}\rho_{+}^4(0)-V_{-}\gamma_{-}\rho_{-}^4(0)\Bigr)
\,\, \biggr] \,,
 \notag
\end{align}
and can finally be rearranged as
\begin{align}\label{B-coefficient-final-expression}
B = \frac{4\pi^3\alpha^3\gamma_{-}}{\kappa_{5}} &  \biggl[ \,\, \frac{\kappa_{5}}{2}(V_{-}-rV_{+})\bar{\rho}^4 -6(1-r)H^2\bar{\rho}^2 + \frac{\kappa_{5}}{\gamma_{-}}\bar{\rho}^4\bar{\sigma}S_{1}
\\
& -\frac{\kappa_{5}}{2}\Bigl( V_{-}\rho_{-}^4(0) -r V_{+}\rho_{+}^4(0)\Bigr)+\Bigl( \rho_{-}^2(0)-r\rho_{+}^{2}(0) \Bigr)H^2 \,\,
\biggr]  \, ,
\notag
\end{align}
where, for convenience, we defined the following ratio of integration constants $r\equiv \frac{\gamma_{+}}{\gamma_{-}}$.
As it will become clear below, $r$ is related to the ratio of two asymptotic values of the radius of the extra dimension during the transition.

As a main difference, compared to the case of Coleman and de Luccia, the bounce depends now not only on $\bar{\rho}$, but also on $\bar{\sigma}$. This means the extremisation of $B$ should somehow be performed with respect to 
two variables rather than one, unless one finds a way of relating them. Additionally, the bounce now also depends on the explicit expressions of the true and false vacuum solutions $\rho_{\pm}(0)$. In turn these depend, as can be seen from the equations appearing in the previous subsection, on various integration constants that should be added to the ratio $r$ already appearing in $B$.

It is thus clear that, even in such a slightly more complicated scenario, proceeding with the CdL approach would involve determining a large set of extra conditions that would allow to fix all these new degrees of freedom. Instead, in the next subsection we continue our analysis by requiring the Israel junction conditions to the case of a compact dimension. These will naturally address the above problem by imposing constraints on the degrees of freedom involved in our setting.

\subsection{Junction conditions with compact dimension}

As demonstrated in the non-compact case of Coleman and de Luccia in subsection~\ref{subsec:israel-CdL}, the Israel junction conditions consist of two parts. The first one is the continuity of the induced metrics across the thin wall, namely
\begin{align}
\bar\rho\equiv\rho(\bar\xi+0)=\rho(\bar\xi-0)\,, \quad
\bar\sig\equiv\sig(\bar\xi+0)=\sig(\bar\xi-0)\,.
\end{align}
The second part is the jumping conditions for the derivatives $\dot\rho,\dot\sig$. They can be obtained
by extracting the second total derivatives in the Ricci tensors for the metric $\tilde g$, which read
\begin{align}
R[\tilde g]_{ij}&=\frac{1}{2H^2}\left(-4H^2-\frac{\rho\dot\rho\dot\sig}{\sig}-2\dot\rho^2-\frac{1}{2}\frac{\mathrm{d}^2\rho^2}{\mathrm{d}\xi^2}\right)g_{ij}\,, \\
R[\tilde g]_{\chi\chi}&=-2\dot\rho^2\sig^2+\rho^2\dot\sig^2-\rho\sig\dot\rho\dot\sig-\frac{1}{2}\frac{\mathrm{d}^2(\rho^2\sig^2)}{\mathrm{d}\xi^2}\,.
\end{align}
Replacing the components of the Ricci tensor with $R_{\mu\nu}=\kappa_5(2g_{\mu\nu}V/3+\pd_\mu\phi\pd_\nu\phi)$ that comes from the Einstein equation, and noting also that $\phi$ depends only on $\xi$, we obtain
\begin{align}
\frac{2}{3}\kappa_5\rho^2V&=-4H^2-\frac{\rho\dot\rho\dot\sig}{\sig}-2\dot\rho^2-\frac{1}{2}\frac{\mathrm{d}^2\rho^2}{\mathrm{d}\xi^2}\,, \\
\frac{2}{3}\kappa_5\rho^2\sig^2V&=-2\dot\rho^2\sig^2+\rho^2\dot\sig^2-\rho\sig\dot\rho\dot\sig-\frac{1}{2}\frac{\mathrm{d}^2(\rho^2\sig^2)}{\mathrm{d}\xi^2}\,.
\end{align}
Integrating each over $\xi$ in an infinitesimal region $[\bar\xi-\de,\bar\xi+\de]$, we obtain
\begin{align}
-\frac{1}{2\bar\rho^2}\frac{\mathrm{d}\rho^2}{\mathrm{d}\xi}\Big|_{\xi=\bar\xi+\de}+\frac{1}{2\bar\rho^2}\frac{\mathrm{d}\rho^2}{\mathrm{d}\xi}\Big|_{\xi=\bar\xi-\de}&=\frac{2}{3}\kappa_5\int_{\bar\xi-\de}^{\bar\xi+\de}\!\mathrm{d}\xi\,V \,, \\
-\frac{1}{2\bar\rho^2\bar\sig^2}\frac{\mathrm{d}(\rho^2\sig^2)}{\mathrm{d}\xi}\Big|_{\xi=\bar\xi+\de}+\frac{1}{2\bar\rho^2\bar\sig^2}\frac{\mathrm{d}(\rho^2\sig^2)}{\mathrm{d}\xi}\Big|_{\xi=\bar\xi-\de}&=\frac{2}{3}\kappa_5\int_{\bar\xi-\de}^{\bar\xi+\de}\!\mathrm{d}\xi\,V \,.
\end{align}
Taking their difference gives the continuity of $\dot\sig$:
\begin{align}
\dot\sig(\bar\xi+0)=\dot\sig(\bar\xi-0)\,,
\end{align}
while the discontinuity of $\dot\rho$ remains the same as in the non-compact case \eqref{jump-d}. 

In summary, the Israel junction conditions are given by
\begin{equation}\label{Israel-junction-conditions}
\begin{aligned}
\rho_{-}(\bar{\xi}) &= \rho_{+}(\bar{\xi}) \, ,
\\
\frac{\kappa_{5}}{3}\bar{\rho}S_{1} &= \dot\rho_{-}(\bar{\xi})-\dot\rho_{+}(\bar{\xi}) \, ,
\\
\sigma_{-}(\bar{\xi}) &= \sigma_{+}(\bar{\xi}) \, ,
\\
\dot\sigma_{-}(\bar{\xi}) &= \dot\sigma_{+}(\bar{\xi}) \,, 
\end{aligned}
\end{equation}
where we used \eqref{int2V=S1}.

\section{Two explicit examples of vacuum transitions}\label{sec:examples}

As in section \ref{sec:theories-with-O(d)-symmetry}, we shall consider the two cases  $V_{+}=\epsilon,\,\, V_{-}=0$ and $V_{+}=0, \,\, V_{-}=-\epsilon$, respectively corresponding to dS $\rightarrow$ Minkowski and Minkowski $\rightarrow$ AdS vacuum transitions.

\subsection{dS $\rightarrow$ Minkowski} 
This scenario corresponds to $V_{+}=\epsilon$ and $V_{-}=0$, namely \textbf{dS} and \textbf{Mink} cases in \eqref{vacuum-solution-4-cases}, and the functions $\rho,\sigma$ take the following explicit forms
\begin{equation}\label{Vacuum-solution-V+=epsilon}
\begin{aligned}
&
\begin{cases}
\rho_{-}(\xi) = \sqrt{2}H\xi
\\
\\
\sigma_{-}(\xi) = \gamma_{-}\xi^{-1}
\end{cases}
\qquad\qquad \qquad \qquad \qquad \qquad \qquad \qquad \qquad  \text{for} \qquad 0\leq\xi\leq\bar{\xi}
\\
\\
& 
\begin{cases}
\rho_{+}(\xi) = \sqrt{\frac{6H^2}{\kappa_{5}\epsilon}+\frac{6\sqrt{H^4+\tfrac{1}{9}\kappa_{5}\epsilon\beta_{+}}}{\kappa_{5}\epsilon}\,\, \sin\biggl[ \sqrt{\frac{2\kappa_{5}\epsilon}{3}}\xi-\xi_{0}^{+} \biggr]} 
\\
\\
\sigma_{+}(\xi) = \gamma_{+} \,\,  \frac{\sqrt{\frac{\kappa_{5}\epsilon}{6}}\sqrt{H^4+\tfrac{1}{9}\kappa_{5}\epsilon\beta_{+}} \,\, \cos\biggl[ \sqrt{\frac{2\kappa_{5}\epsilon}{3}}\xi-\xi_{0}^{+} \biggr]}{H^2+\sqrt{H^4+\tfrac{1}{9}\kappa_{5}\epsilon\beta_{+}}\,\, \sin\biggl[ \sqrt{\frac{2\kappa_{5}\epsilon}{3}}\xi-\xi_{0}^{+} \biggr]}
\end{cases}
\qquad \,\,\, \text{for} \qquad \xi \geq \bar{\xi}
\end{aligned}
\end{equation}
where the solution inside the bubble is obtained from \eqref{Vacuum-solution-V=0} by requiring that $\rho(0)=0$, which enforces $\beta_{-}=0=\xi_{0}^{-}$, and the one outside is simply \eqref{Vacuum-solution-Vc>0}, with $-\frac{9H^4}{\kappa_{5}\epsilon} < \beta_{+} \leq 0$. 
Note that the solution inside the bubble describes flat space $\mathbb{R}^4\times S^1$ with radius:
\begin{equation}\label{R0}
R_0=\sqrt{2}H\gamma_- \,.
\end{equation}

To continue we impose $\rho_{-}(\bar{\xi})=\rho_{+}(\bar{\xi})$, namely the first continuity condition in \eqref{Israel-junction-conditions}, which fixes the integration constant $\xi_{0}^{+}$ to be of the form
\begin{equation}\label{integration-constant-dS-to-Mink}
\xi_{0}^{+}=\sqrt{\frac{2\kappa_{5}\epsilon}{3}}\bar{\xi}-\arcsin{\biggl[ \frac{H^2(\kappa_{5}\epsilon\bar{\xi}^2-3)}{\sqrt{9H^4+\kappa_{5}\epsilon\beta_{+}}} \biggr]} \, .
\end{equation}
Note that $\beta_{+}$ is required to satisfy $-\frac{9H^4}{\kappa_{5}\epsilon} < \beta_{+} \le 0$, which makes $\xi_{0}^{+}$ well-defined in the whole range. To understand the geometry of the solution outside the bubble we consider the gravity decoupled limit $\kappa_5\to 0$. Then $\xi_{0}^{+}\simeq \sqrt{\frac{2\kappa_{5}\epsilon}{3}}\bar{\xi} +{\pi\over 2}$ and \eqref{Vacuum-solution-V+=epsilon} yields:
\begin{equation}\label{approx-solution-outside}
\rho_+(\xi)\sim\sqrt{12\over\kappa_5\epsilon}H\left|\sin\sqrt{\frac{\kappa_{5}\epsilon}{6}}(\xi-\bar\xi)\right|
\quad;\quad
\rho_+\sigma_+(\xi)\sim \sqrt{2}H\gamma_+\left|\cos\sqrt{\frac{\kappa_{5}\epsilon}{6}}(\xi-\bar\xi)\right|\,,
\end{equation}
describing locally the product of a sphere $S^3$ of radius $\sqrt{6/\kappa_5\epsilon}$ and a circle of oscillating radius around the value $\sqrt{2}H\gamma_+$.

The connection between $\bar{\rho}$ and $\rho_{-}(\bar{\xi})$ provides then the identification
\begin{equation}\label{relation-xi-rhobar}
\bar{\xi} = \frac{\bar{\rho}}{\sqrt{2}H} \, ,
\end{equation}
while the relation between $\bar{\rho}$ and $\rho_{+}(\bar{\xi})$ leads to the following trigonometric relations
\begin{equation}\label{trigonometric-relations}
\begin{aligned}
\sqrt{9H^4+\kappa_{5}\epsilon\beta_{+}} \sin{\Bigl[ \sqrt{\frac{2\kappa_{5}\epsilon}{3}}\bar{\xi}-\xi_{0}^{+} \Bigr]} & = \frac{1}{2}\kappa_{5}\epsilon\bar{\rho}^2-3H^2 \,,
\\
\sqrt{9H^4+\kappa_{5}\epsilon\beta_{+}} \cos{\Bigl[ \sqrt{\frac{2\kappa_{5}\epsilon}{3}}\bar{\xi}-\xi_{0}^{+} \Bigr]} &= \sqrt{\kappa_{5}\epsilon\beta_{+}+3H^2\kappa_{5}\epsilon\bar{\rho}^2-\frac{1}{4}\kappa_{5}^2\epsilon^2\bar{\rho}^4} \, .
\end{aligned}
\end{equation}
The identities \eqref{relation-xi-rhobar} and \eqref{trigonometric-relations} can now be exploited to simplify all the subsequent constraints. The jumping condition on $\dot{\rho}$, namely the second equation in \eqref{Israel-junction-conditions}, can be translated into an expression for the integration constant $\beta_{+}$
\begin{equation}\label{expression-for-beta+}
\beta_{+}=\frac{1}{4}\left(\kappa_{5}\epsilon+\frac{2}{3}\kappa_{5}^2S_{1}^2\right)\bar{\rho}^4-\sqrt{2}H\kappa_{5}S_{1}\bar{\rho}^3 \,,
\end{equation}
while continuity of $\sigma,\dot\sigma$, namely the third and fourth conditions in \eqref{Israel-junction-conditions}, respectively translate into the following two expressions for the ratio of integration constants $\gamma_{\pm}$,
\begin{equation}\label{expressions-for-r}
\begin{aligned}
\frac{1}{r}\equiv \frac{\gamma_{-}}{\gamma_{+}}&=\frac{1}{\sqrt{2}H\bar{\rho}}\sqrt{\frac{2}{3}\beta_{+}+2H^2\bar{\rho}^2-\frac{1}{6}\kappa_{5}\epsilon\bar{\rho}^4} \, ,
\\
\frac{1}{r}\equiv \frac{\gamma_{-}}{\gamma_{+}}&= 1+\frac{2\beta_{+}}{3H^2\bar{\rho}^2} \, .
\end{aligned}
\end{equation}
Consistency between the three conditions \eqref{expression-for-beta+} and \eqref{expressions-for-r} leads to the following solutions
\begin{align}
\textbf{dS1.} \qquad
\frac{1}{r}&=1-\frac{\kappa_{5}S_{1}}{3\sqrt{2}H}\bar{\rho}
\qquad \quad
\beta_{+}=-\frac{\sqrt{2}H\kappa_{5}S_{1}}{4}\bar{\rho}^3
\qquad \quad
\bar{\rho} = \frac{9\sqrt{2}H S_{1}}{3\epsilon + 2\kappa_{5} S_{1}^2}
\label{sol1}\\
\notag \\
\textbf{dS2.}\qquad 
\frac{1}{r}&=-1+\frac{\kappa_{5}S_{1}}{3\sqrt{2}H}\bar{\rho}
\qquad  \qquad \qquad 
\beta_{+}=-3H^2\bar{\rho}^2\Bigl( 1-\frac{\kappa_{5}S_{1}}{6\sqrt{2}H}\bar{\rho} \Bigr) 
\label{sol2}\\
& \qquad  \bar{\rho}=\frac{15\sqrt{2}H\kappa_{5}S_{1}\pm 3\sqrt{6}H\sqrt{3\kappa_{5}^2S_{1}^2-8\kappa_{5}\epsilon}}{4\kappa_{5}^2S_{1}^2+6\kappa_{5}\epsilon}
\notag
\end{align}

\vspace{5mm}
\noindent
\textbf{dS1.} \qquad The first solution exhibits an expression for $\bar{\rho}$ which has the same structure as the one found in the CdL case without compact directions \eqref{CdL-V+=epsiolon-V-=0-solution}. Substituting into $r$ and $\beta_{+}$ we obtain
\begin{equation}\label{Sol1.}
\frac{1}{r}=1-\frac{3\kappa_{5}S_{1}^2}{2\kappa_{5}S_{1}^2+3\epsilon} \, ,
\qquad
\beta_{+}=-\frac{729H^4\kappa_{5}S_{1}^4}{(2\kappa_{5}S_{1}^2+3\epsilon)^3} \, ,
\qquad
\bar{\rho} = \frac{9\sqrt{2}H S_{1}}{3\epsilon + 2\kappa_{5} S_{1}^2} \, ,
\end{equation}
which makes it evident that $\bar{\rho}\geq 0$ and $\frac{1}{r} \simeq 1$ up to gravitational corrections. This also allows to compare $\beta_{+}$ with its allowed range $-\frac{9H^4}{\kappa_{5}\epsilon} < \beta_{+} \le 0$. It is clear that $\beta_{+}$ in \eqref{Sol1.} is negative, hence we only need to compare with the lower bound of the domain. This translates into the condition
\begin{equation}
-\frac{729H^4\kappa_{5}S_{1}^4}{(2\kappa_{5}S_{1}^2+3\epsilon)^3}+\frac{9H^4}{\kappa_{5}\epsilon} > 0 \,,
\end{equation}
which can be recast in the form
\begin{equation}
\frac{9H^4(\kappa_{5}S_{1}^2-3\epsilon)^2(8\kappa_{5}S_{1}^2+3\epsilon)}{\kappa_{5}\epsilon(2\kappa_{5}S_{1}^2+3\epsilon)^3} > 0 \, .
\end{equation}
This is clearly respected for any $S_{1}$ since $\epsilon$ and $\kappa_{5}$ are positive quantities. Solution \eqref{Sol1.} is therefore well-defined, as it also exhibits a consistent limit for the decoupling of gravity. Expanding the quantities in \eqref{sol1} around $\kappa_{5}\rightarrow 0$ one obtains
\begin{equation}
\bar{\rho}=\frac{3\sqrt{2}HS_{1}}{\epsilon} \, , \qquad
{r}=1 \, ,
\qquad
\beta_{+}=0 \, .
\end{equation}

At this point, we just need to explicitly compute the bounce action $B$, providing the exponent of the transition probability, using the expression \eqref{B-coefficient-final-expression} and substituting in it the various relations found above. One gets
\begin{align}
B=&\frac{\pi^3\gamma_{-}}{2\epsilon H^2\kappa_{5}^2(3\sqrt{2}H\!-\!\kappa_{5}S_{1}\bar{\rho})}
\biggl[ -\! \bar{\rho}^4 2\epsilon\kappa_{5}^2(2\kappa_{5}S_{1}^2+3\epsilon) \!+\! \bar{\rho}^3 3\sqrt{2}\epsilon H\kappa_{5}^2 S_{1}(7 \!+\! \cos{2\xi_{0}^{+}}) 
\\
& \qquad \qquad \qquad \!-\! 60H^2\sqrt{36H^4 \!-\! \sqrt{2}H\epsilon\kappa_{5}^2S_{1}\bar{\rho}^3}\sin{\xi_{0}^{+}} \!+\! 36H^4(7-3\cos{2\xi_{0}^{+}})
\biggr] \, ,
\notag
\end{align}
with 
\begin{equation}
\begin{aligned}
\xi_{0}^{+}&=\sqrt{\frac{\epsilon\kappa_{5}}{3H^2}}\bar{\rho}+\arcsin{\biggl[ \frac{6H^2-\epsilon\kappa_{5}\bar{\rho}^2}{\sqrt{36H^4-\sqrt{2}H\epsilon\kappa_{5}^2S_{1}\bar{\rho}^3}} \biggr]} \equiv a + \arcsin{(b)} \, ,
\\
\sin{(\xi_{0}^{+})}&=b \cos{(a)}+\sqrt{1-b^2}\sin{(a)} \, .
\end{aligned}
\end{equation}
Substituting $\bar{\rho}$ from \eqref{sol1} and exploiting the dimensionless quantity $\hat{\kappa}_{5}$, introduced in \eqref{dimensionless-kappa} at the end of subsection \ref{subsec:CdL}, one finds the following expression for $B$
\begin{equation}
B(\gamma_{-},H,\kappa_{5},S_{1},\epsilon) = 45\frac{\pi^3 R_{0} S_{1}^4}{\epsilon^3} \mathcal{B}(\hat{\kappa}_{5}) \,,
\end{equation}
with $\mathcal{B}(\hat{\kappa}_{5})$ dimensionless and allowing the following series expansion in the gravity decoupling limit $\hat{\kappa}_{5}\rightarrow 0$
\begin{equation}
B \simeq 45 \frac{\pi^3 R_{0} S_{1}^4}{\epsilon^3} \Bigl( 1 - \hat{\kappa}_{5} +\frac{7}{4}\hat{\kappa}_{5}^2 + \mathcal{O}(\hat{\kappa}_{5}^3) \Bigr) \, .
\end{equation}
Another important property of the above bounce action is that it only depends on $H$ and $\gamma_{-}$ via the combination $R_{0}\equiv \sqrt{2}\gamma_{-}H$ which is the radius of the compact dimension in the true vacuum (inside the bubble), see \eqref{R0}. Its value at the origin $R_{0} \equiv (\rho \sigma)|_{\xi=0}$ also coincides with the radius at $\xi=\bar{\xi}$, namely $R_{\bar{\xi}}\equiv ( \rho \sigma )|_{\xi=\bar{\xi}} = R_{0}$.
Note that in the gravity decoupled limit, $B$ is proportional to the only dimensionless quantity of $S_1,\epsilon, R_0$ which is linear in $R_0$ and analytic in $S_1$, appearing in the lowest integer power. This is to be contrasted with the dimensionless ratio $S_1^5/\epsilon^4$ obtained in the non compact case \eqref{Bnoncompact}. This is obtained by replacing $R_0$ by $S_1/\epsilon$ which has dimension of length, up to a numerical constant. On the other hand, in the limit of vanishing radius, one recovers the 4-dimensional quantity $S_1^4/\epsilon^3$ upon rescaling $S_1$ and $\epsilon$ by $1/R_0$ as dictated by standard dimensional reduction.

We can finally compare, see figure \ref{fig:dS}, the explicit plots of the dimensionless functions $\mathcal{B}(\hat{\kappa}_{d})$, in the range $\hat{\kappa}_{d} \in [0,1]$, providing the gravitational corrections to the dS$\rightarrow$Minkowski transitions discussed for the non-compact case of $d=4,5$ in subsection \ref{subsec:CdL} and for $d=5$ with one compact dimension above. Note that, contrary to what one may naively expect, the curve for the $d=5$ case in the presence of a compact dimension does not lie between the curves for the non-compact $d=4$ and $d=5$ cases. Moreover, its radius dependence is trivial and does not interpolate between  the two non-compact curves in the limits $R_{0}\rightarrow 0$ and $R_{0} \rightarrow \infty$. This result stems from the structure of the Euclidean equations of motion in the compact case \eqref{EeqS1-3}, which do not reduce to the non-compact ones \eqref{Eeqd2} for constant $\sigma(\xi)$. 
In fact, the equations of motion \eqref{EeqS1-3} turn out to be incompatible with solutions where $\sigma(\xi)$ is a constant, even asymptotically. This can be seen by noting that \eqref{sigeq} implies that, if $\sigma(\xi)$ is constant, $\rho(\xi)$ grows exponentially in $\xi$ with an exponent $\frac{\sigma}{\gamma}$, while addition of eqs \eqref{EeqS1-1} and \eqref{EeqS1-3} gives
\begin{equation}
6\left({\sigma\over\gamma}\right)^2+\kappa_{d} V={3\over\alpha^2\rho^2} \,\, \to \,\, 0 \qquad \text{for} \qquad \xi \,\, \to \,\, \infty \,\, ,
\end{equation}
which is incompatible with constant $\sigma(\xi)$ for positive $V$ (the case of negative $V$ corresponding to a transition between two AdS vacua is more involved). Notice however that in the Lorentzian case the two terms on the left-hand side above come with a relative minus sign and there is no obstruction; one obtains a 5d de Sitter solution where the compact and non-compact dimensions expand exponentially with time~\cite{Anchordoqui:2023etp}. 

In turn, the obstruction arising for solutions with constant $\sigma(\xi)$ can also be regarded as the impossibility of changing the topological structure of the bounce from $O(4)\times U(1)$ to $O(5)$ by simply taking the limit of the $U(1)$-radius to infinity, as one may naively expect. We notice that this limit exists in the case of flat spatial sections, i.e. $S^3$ is replaced by $\mathbb{R}^3$, leading to a locally dS$_5$ or AdS$_5$ solution (depending on the sign of the potential $V$) with one compact space dimension. In this case, there is no topological obstruction but the solution has not finite action.\vspace{-0.2cm}
\begin{figure}[H]
\centering
\includegraphics[scale=0.25]{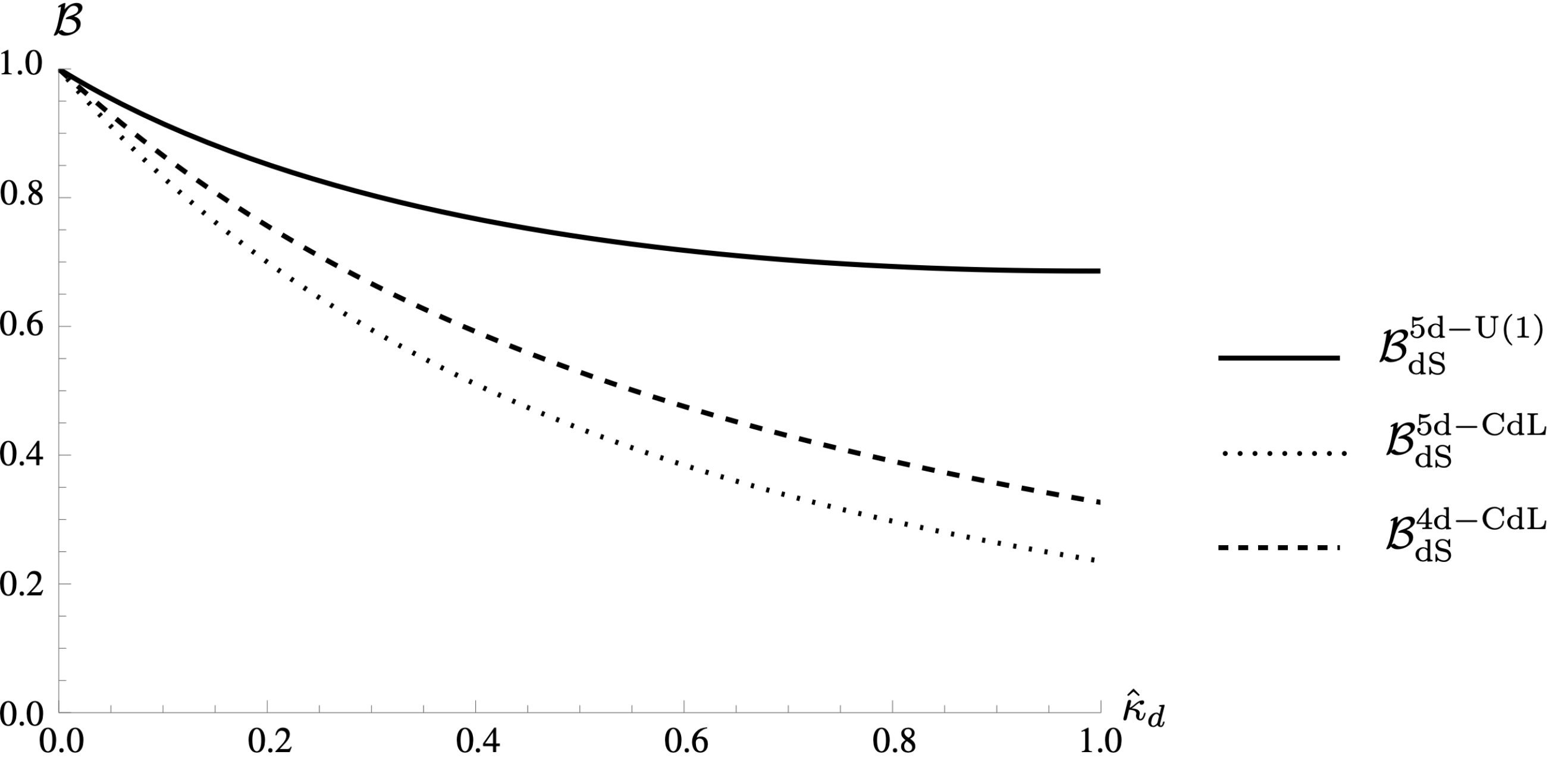}
\caption{Behaviour of the dimensionless function $\mathcal{B}(\hat{\kappa}_{d})$ describing the gravitational corrections in the dS to Minkowski transition, normalised to the result in the absence of gravity, as a function of the dimensionless parameter $\hat\kappa_d$ defined in \eqref{dimensionless-kappa}. The plot compares the results of the $d=4,5$ non-compact case obtained in section~\ref{sec:theories-with-O(d)-symmetry} with those of $d=5$ with one compact dimension studied here.}
\label{fig:dS}
\end{figure}

\noindent
\textbf{dS2.} \qquad The second solution is excluded because it does not have a sensible gravity decoupled limit. Indeed, when
$\kappa_{5}\rightarrow 0$, 
one expects to recover a well-defined expression for $\bar{\rho}$, as well as for the integration constants found above, namely $\frac{1}{r}=1$ and $\beta_{+}=0$. Expanding the quantities \eqref{sol2} in such a limit, one finds a complex value for $\bar\rho$, a negative value for the ratio $r$ and a divergent $\beta_+$ which are all troublesome, leading to
the exclusion of this solution.

\subsection{Minkowski $\rightarrow$ AdS} 
This transition corresponds to the choice $V_{+}=0$ and $V_{-}=-\epsilon$, respectively associated with \textbf{Mink} and \textbf{AdS1}, \textbf{AdS2}, \textbf{AdS3} cases in \eqref{vacuum-solution-4-cases}. We consider them in order below. 

\vspace{2mm}
\noindent
\textbf{AdS1.} \qquad The functions $\rho$ and $\sigma$ take the following form
\begin{equation}
\begin{aligned}
&
\begin{cases}
\rho_{-}(\xi)=\frac{2\sqrt{3}H}{\sqrt{\kappa_{5}\epsilon}} \sinh{\Bigl[ \sqrt{\frac{\kappa_{5}\epsilon}{6}}\xi \Bigr]}
\\
\\
\sigma_{-}(\xi)=\gamma_{-}\sqrt{\frac{\kappa_{5}\epsilon}{6}} \coth{\Bigl[ \sqrt{\frac{\kappa_{5}\epsilon}{6}} \xi \Bigr]}
\end{cases}
\qquad\qquad \,\text{for} \qquad 0\leq\xi\leq\bar{\xi}
\\
\\
& 
\begin{cases}
\rho_{+}(\xi)=\sqrt{2H^2(\xi-\xi_{0}^{+})^2-\frac{\beta_{+}}{3H^2}}
\\
\\
\sigma_{+}(\xi)=-\gamma_{+} \,\, \frac{6H^4(\xi-\xi_{0}^{+})}{\beta_{+} - 6H^4(\xi-\xi_{0}^{+})^2}
\end{cases}
\qquad\qquad \text{for} \qquad \xi \geq \bar{\xi}
\end{aligned}
\end{equation}
The solution inside the bubble is obtained from \eqref{Case3-general-solution} by first  requiring the root $x_{-}$ to vanish. This fixes $\beta_{-}=0$ and makes sure that the solution $\rho_{-}(\xi)$ is allowed to start from $\xi=0$. Demanding then its vanishing at the origin further fixes the integration constant 
\begin{equation}
\xi_{0}^{-}=\log{\sqrt{8}}
\end{equation}
and leads to the above simplified expression for $\rho_{-}(\xi)$. It describes locally a Euclidean AdS$_4$ space of inverse radius $\sqrt{\frac{\kappa_{5}\epsilon}{6}}$ times a circle of radius that stays approximately constant, starting at the origin from $R_0\simeq\sqrt{2}H\gamma_-$ as in \eqref{R0} and increasing slightly due to gravitational corrections (see below).
The solution outside the bubble is simply \eqref{Vacuum-solution-V=0}, with the requirement that $\beta_{+}\le 0$.  It describes asymptotically flat space $\mathbb{R}^4\times S^1$ of radius $R_\infty\simeq\sqrt{2}H\gamma_+$.

We now proceed as in the dS $\rightarrow$ Minkowski transition, by first imposing that $\rho_{-}(\bar{\xi})=\rho_{+}(\bar{\xi})$, namely the first condition in \eqref{Israel-junction-conditions}. This fixes the integration constant $\xi_{0}^{+}$ to 
\begin{equation}
\xi_{0}^{+} = \bar{\xi} \pm \sqrt{\frac{\beta_{+}}{6H^4}+\frac{6}{\kappa_{5}\epsilon}\sinh^2{\left[\sqrt{\frac{\kappa_{5}\epsilon}{6}}\bar{\xi}  \right]}} \,.
\end{equation}
Notice that the above square root further restricts the allowed domain of $\beta_{+}$ as 
\begin{equation}\label{betap-restriction}
-\frac{36H^4} {\kappa_{5}\epsilon} \sinh^2{\left[\sqrt{\frac{\kappa_{5}\epsilon}{6}}\bar{\xi}  \right]} \le \beta_{+}\le 0 \, .
\end{equation}
The connection between $\bar{\rho}$ and $\rho_{-}(\bar{\xi})$ leads to the following identities
\begin{equation}\label{trig-identities2}
\sinh{\left[\sqrt{\frac{\kappa_{5}\epsilon}{6}}\bar{\xi}  \right]} = \bar{\rho} \sqrt{\frac{\kappa_{5}\epsilon}{12H^2}} \, , \qquad \qquad  \cosh{\left[\sqrt{\frac{\kappa_{5}\epsilon}{6}}\bar{\xi}  \right]} = \sqrt{1+ \frac{\kappa_{5}\epsilon}{12H^2}\bar{\rho}^2} \, ,
\end{equation}
while the connection between $\bar{\rho}$ and $\rho_{+}(\bar{\xi})$ allows to identify
\begin{equation}\label{rel-rhoplus}
H(\bar{\xi}-\xi_{0}^{+}) = \sqrt{\frac{1}{2}\bar{\rho}^2+\frac{1}{6H^2}\beta_{+}} \,.
\end{equation}
The relations \eqref{trig-identities2} and \eqref{rel-rhoplus} can now be exploited to simplify all the remaining conditions in \eqref{Israel-junction-conditions}. The jumping condition on $\dot{\rho}$ leads to
\begin{equation}\label{cond1}
\beta_{+}=-3H^2\bar{\rho}^2\left[ 1-\frac{1}{2H^2}\biggl( \sqrt{2H^2+\frac{1}{6}\kappa_{5}\epsilon\bar{\rho}^2}-\frac{1}{3}\kappa_{5}S_{1}\bar{\rho}\biggr)^2
 \right] \,,
\end{equation}
while the continuity conditions on $\sigma,\dot\sigma$ translate into expressions for the ratio of $\gamma_{\pm}$
\begin{equation}\label{cond2}
\begin{aligned}
\frac{1}{r}\equiv \frac{\gamma_{-}}{\gamma_{+}}&=\sqrt{\frac{12H^2\bar{\rho}^2+4\beta_{+}}{12H^2\bar{\rho}^2+\kappa_{5}\epsilon\bar{\rho}^4}}
\\
\frac{1}{r}\equiv \frac{\gamma_{-}}{\gamma_{+}}&=1+\frac{2\beta_{+}}{3H^2\bar{\rho}^2} \, .
\end{aligned}
\end{equation}
Consistency between the three conditions \eqref{cond1} and \eqref{cond2} is now more involved than the previous case, but still allows to fix $r,\beta_{+},\bar{\rho}$. After some manipulations one finds a fraction of polynomials in $\bar{\rho}^2$, with numerator and denominator respectively of fifth and second order in $\bar{\rho}^2$, the roots of which provide the desired solution. This takes, up to overall prefactors, the following form
\begin{align}\label{fraction-of-polynomials}
\frac{N}{D^2}&= \frac{P_{2}P_{3}}{D^2} \equiv 0  \qquad \qquad \text{with}
\\
P_{2}&= \bar{\rho}^4 \epsilon\kappa_{5}(2\kappa_{5}S_{1}^2-3\epsilon)^2+\bar{\rho}^2 12H^2(4\kappa_{5}^2S_{1}^4-24\epsilon\kappa_{5}S_{1}^2+9\epsilon^2)-1944H^4S_{1}^2 \,,
\notag \\
P_{3}&=\bar{\rho}^6\epsilon\kappa_{5}^3(2\kappa_{5}S_{1}^2-3\epsilon)^2 +\bar{\rho}^4 12H^2\kappa_{5}^2(4\kappa_{5}^2 S_{1}^4-36\epsilon\kappa_{5}S_{1}^2+27\epsilon^2) 
\notag \\
&~\quad + \bar{\rho}^2 216H^4 \kappa_{5}(18\epsilon-17\kappa_{5}S_{1}^2) + 15552H^6 \,,
\notag \\
D &= \bar{\rho}^4 \epsilon\kappa_{5}^2(2\kappa_{5}S_{1}^2+3\epsilon) + \bar{\rho}^2 6H^2\kappa_{5}(4\kappa_{5}S_{1}^2+9\epsilon) + 162H^4 \,.
\notag 
\end{align}
Since the denominator $D$ is a non-vanishing quantity, the next step is finding the roots of the numerator which are compatible with our needs. We start by analysing the third order polynomial, whose discriminant reads, up to overall positive factors 
\begin{equation}
\Delta_{P_{3}} \propto 12\kappa_{5}^2S_{1}^4 +70 \epsilon\kappa_{5} S_{1}^2 -81\epsilon^2 \,.
\end{equation}
It is then clear that its sign and the nature of its roots depend on the interplay between $S_{1}$ and $\epsilon$, and one should in principle look for allowed roots taking this into account. To avoid this certainly not-easy route, one can look instead at the gravity-decoupled limit, $\kappa_{5}\rightarrow 0$, of the above equation, which should be well defined and still provide a solution for $\bar{\rho}$. In this limit, the fraction in \eqref{fraction-of-polynomials} takes the following form
\begin{equation}\label{limit}
\frac{N}{D^2} \simeq \frac{4(18H^2S_{1}^2 - \epsilon^2\bar{\rho}^2)}{81H^4} + \mathcal{O}(\kappa_{5}) \equiv 0 \qquad \Rightarrow \qquad \bar{\rho}=\frac{3\sqrt{2}H|S_{1}|}{\epsilon}
\end{equation}
from which a single solution for $\bar{\rho}$ can be extracted depending on the sign of $S_{1}$. The root of the above polynomials leading to such a limit should thus be taken as the meaningful solution and one can immediately recognise that this should come from $P_{2}$, whose always-real roots read
\begin{equation}
\bar{\rho}^2\!=\! \frac{6H^2\left(-4\kappa_{5}^2S_{1}^4\!+\!24\epsilon\kappa_{5}S_{1}^2\!-\!9\epsilon^2 \pm (2\kappa_{5}S_{1}^2\!+\!3\epsilon)\sqrt{4\kappa_{5}^2S_{1}^4\!-\!6\epsilon\kappa_{5}S_{1}^2\!+\!9\epsilon^2}  \right)}{\epsilon \kappa_{5}(2\kappa_{5}S_{1}^2\!-\!3\epsilon)^2} \, .
\end{equation}
The square root in the numerator of the two solutions is positive for any $S_{1}\in \mathbb{R}$ and since $\bar{\rho}^2$ should be a positive quantity with finite gravity decoupled limit, only the solution with the plus sign is allowed. This correctly recovers the limit \eqref{limit} and is always positive. Moreover the zeros of the denominator, when $S_{1}\neq \pm \sqrt{\frac{3\epsilon}{2\kappa_{5}}}$, are beyond the validity of the effective field theory ${\hat\kappa}_5<1$.
This solution leads to the following expressions for the desired quantities
\begin{equation}\label{desired-quantities}
\begin{aligned}
\bar{\rho}\!&=\!\sqrt{ \frac{6H^2\left(-4\kappa_{5}^2S_{1}^4\!+\!24\epsilon\kappa_{5}S_{1}^2\!-\!9\epsilon^2 + (2\kappa_{5}S_{1}^2\!+\!3\epsilon)C  \right)}{\epsilon \kappa_{5}(2\kappa_{5}S_{1}^2\!-\!3\epsilon)^2}} \,,
\\
\beta_{+}\!&=\! \frac{6H^4\left(8\kappa_{5}^3S_{1}^6\!-\!2\kappa_{5}^2 S_{1}^4(9\epsilon\!+\!2C)\!-\!3\epsilon\kappa_{5}S_{1}^2(9\epsilon\!-\!2C)\!+\!9\epsilon^2(3\epsilon\!-\!C) \right)}{\kappa_{5}\epsilon^2(2\kappa_{5}S_{1}^2\!-\!3\epsilon)^2} \,,
\\
\frac{1}{r}\!&=\!\frac{-2\kappa_{5}S_{1}^2+C}{3\epsilon} \, ,
\end{aligned}
\end{equation}
where to shorten the equations we defined $C\equiv \sqrt{4\kappa_{5}^2S_{1}^4\!-\!6\epsilon\kappa_{5}S_{1}^2\!+\!9\epsilon^2}>0$. One should now compare the expression for $\beta_{+}$ with its allowed range
\begin{equation}
-3H^2\bar{\rho}^2\le \beta_{+}\le 0 \, ,
\end{equation}
obtained by combining \eqref{betap-restriction} and \eqref{trig-identities2}, which is always satisfied. Finally, one should check positivity of $\frac{1}{r}$, related to the ratio of the two asymptotic values of the radius of the compact dimension. This imposes
\begin{equation}\label{S1-constraint-AdS}
|S_{1}| < \sqrt{\frac{3\epsilon}{2\kappa_{5}}} \,, 
\end{equation}
which is also consistent with the excluded values of $S_{1}$ from the above denominators.
Once again, in the limit $\kappa_{5}\rightarrow 0$ where gravity decouples, one obtains
\begin{equation}
\bar{\rho} = \frac{3\sqrt{2}H|S_{1}|}{\epsilon} \,, \qquad \beta_{+}=0 \,, \qquad
r=1 \, .
\end{equation}

We can now compute the bounce action $B$ using the expression \eqref{B-coefficient-final-expression}
\begin{equation}\label{B-AdS}
\begin{aligned}
B&=\frac{\pi^3\gamma_{-}}{3\sqrt{2}H^3\kappa_{5}} \biggl[ 2r(\beta_{+}-6H^4(\xi_{0}^{+})^2+18H^2\bar{\rho}^2)
\\
& \qquad \qquad \qquad +\bar{\rho}^2\Bigl(-36H^2-3\kappa_{5}\epsilon\bar{\rho}^2+\kappa_{5}S_{1}\bar{\rho}\sqrt{72H^2+6\epsilon\kappa_{5}\bar{\rho}^2} \Bigr) \biggr]
\end{aligned}
\end{equation}
with $r,\beta_{+},\bar{\rho}$ given in \eqref{desired-quantities} and
\begin{equation}
\xi_{0}^{+}=\sqrt{\frac{6}{\kappa_{5}\epsilon}}\arcsinh{\left( \frac{\sqrt{\epsilon\kappa_{5}}\bar{\rho}}{2\sqrt{3}H} \right)}
-\sqrt{\frac{\beta_{+}}{6H^4}+\frac{\bar{\rho}^2}{2H^2}} \,.
\end{equation}
Substituting the above quantities in \eqref{B-AdS} and exploiting $\hat{\kappa}_{5}$ introduced in \eqref{dimensionless-kappa}, one finds an expression for $B$ of the form encountered above
\begin{equation}
B(H,\gamma_{-},\kappa_{5},S_{1},\epsilon) = 45\frac{\pi^3 R_{0} S_{1}^4}{\epsilon^3} \mathcal{B}(\hat{\kappa}_{5}) \,,
\end{equation}
where once again the dependence on $H$ and $\gamma_{-}$ appears only via the radius of the compact dimension at $\xi=0$, namely $R_{0} \equiv ( \rho \sigma )|_{\xi=0} = \sqrt{2}\gamma_{-}H$. Note that, contrary to the dS$\rightarrow$Minkowski transition, the radius of the circle is now not constant within the bubble, but as mentioned above, it receives gravitational corrections as $\xi$ varies, so that its respective values at the origin, on the wall and at infinity read
\begin{equation}
\begin{aligned}
R_{0}&\equiv ( \rho\sigma ) |_{\xi=0} = \sqrt{2}\gamma_{-}H  \,,
\\
R_{\bar{\xi}}&\equiv ( \rho\sigma )|_{\xi=\bar{\xi}} = R_0\biggl( 1+\frac{\kappa_{5}\epsilon}{12H^2}\bar{\rho}^2 \biggr)
\simeq R_0(1+ {3\over 2}{\hat\kappa}_5+\dots) \,,
\\ 
R_{\infty}&\equiv ( \rho\sigma )|_{\xi=\bar{\xi}} = R_0 r=R_0(1+{\hat\kappa}_5+\dots)\,.
\end{aligned}
\end{equation}
Another important difference, in comparison to the dS$\rightarrow$Minkowski case, is that now the form of $\mathcal{B}(\hat{\kappa}_{5})$ depends on whether one chooses $S_{1}\geq 0$ or $S_{1}\le 0$, corresponding to positive or negative brane tension. It turns out that the second choice is excluded because it leads to a value for $B$ which is always negative. In light of the condition \eqref{S1-constraint-AdS}, one finally has the allowed range
\begin{equation}\label{S1-allowed-final-AdS}
0 \le S_{1} < \sqrt{\frac{3\epsilon}{2\kappa_{5}}} \,,
\end{equation}
which leads to
a bounce action with the following expansion in the gravity decoupling limit
\begin{equation}
B \simeq 45 \frac{\pi^3 R_{0} S_{1}^4}{\epsilon^3} \Bigl( 1 + 2\hat{\kappa}_{5} +\frac{11}{4}\hat{\kappa}_{5}^2 + \mathcal{O}(\hat{\kappa}_{5}^3) \Bigr)\,.
\end{equation}
We finally notice that, when introducing the dimensionless variable $\hat{\kappa}_{5}$, the allowed range \eqref{S1-allowed-final-AdS} of $S_{1}$ translates into $0 \le \hat{\kappa}_{5} \le \frac{3}{2}$, which does not restrict the domain of interest $0\le \hat{\kappa}_{5} \le 1$.
Hence, we can once again compare, see figure \ref{fig:AdS}, the plots for $\mathcal{B}(\hat{\kappa}_{d})$, in the range $\hat{\kappa}_{d} \in [0,1]$, for the Minkowski to AdS transition discussed in subsection \ref{subsec:CdL} for $d=4,5$ and above for $d=5$ with a compact dimension. Note that, as already observed in the dS to Minkowski transition, the curve for the $d=5$ case with one compact dimension does not lie between the $d=4$ and $d=5$ non-compact curves, and does not interpolate between the limits $R_{0}\rightarrow 0$ and $R_{0} \rightarrow \infty$; see discussion above figure \ref{fig:dS} and conclusions.
\begin{figure}[H]
\centering
\includegraphics[scale=0.25]{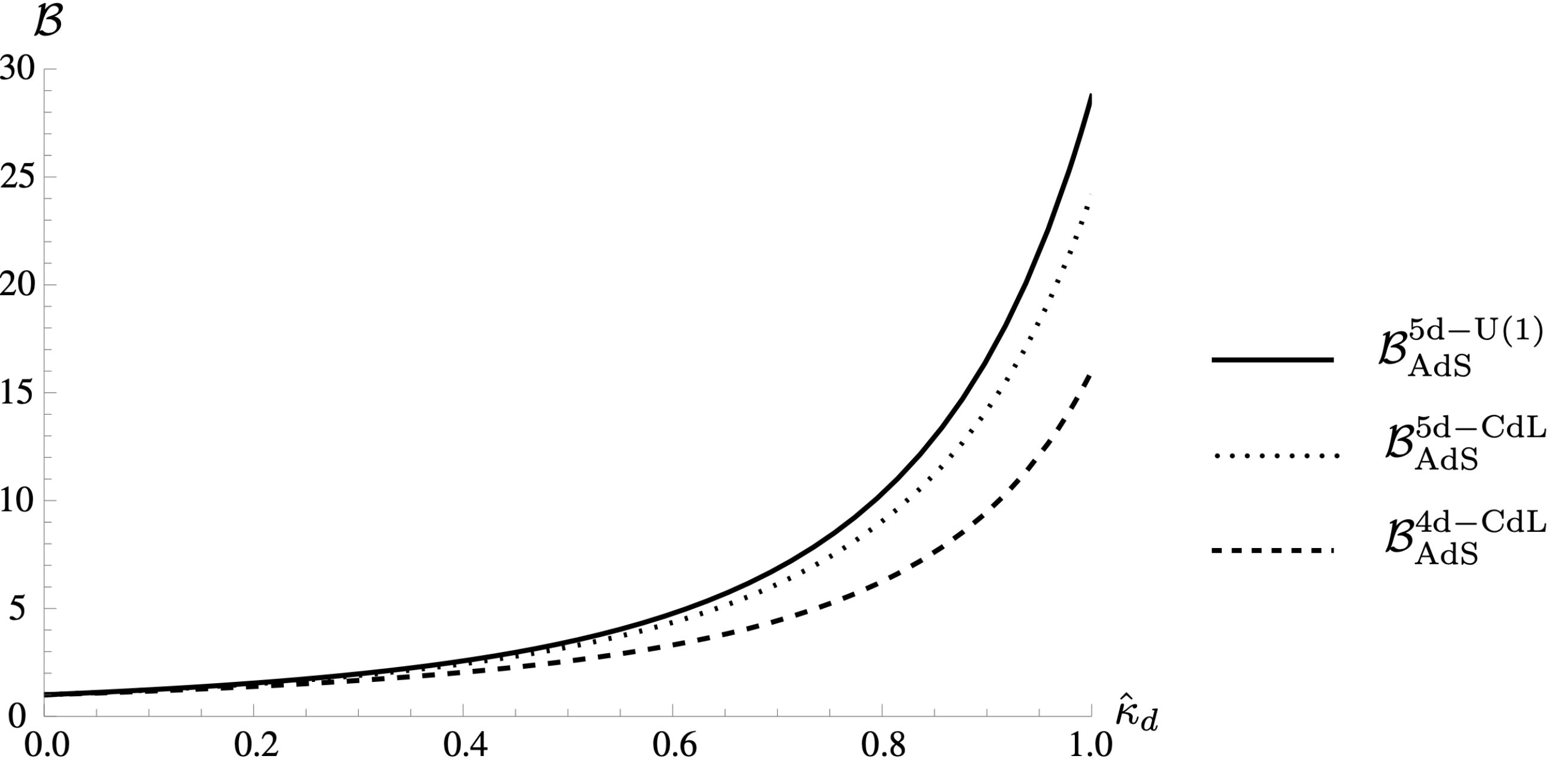}
\caption{Behaviour of the dimensionless function $\mathcal{B}(\hat{\kappa}_{d})$ describing the gravitational corrections in the Minkowski to AdS transition. The plot compares the results of the $d=4,5$ non-compact case obtained in section~\ref{sec:theories-with-O(d)-symmetry} with those of $d=5$ with one compact dimension studied here.}
\label{fig:AdS}
\end{figure}

\vspace{5mm}
\noindent
\textbf{AdS2.} \qquad The functions $\rho$ and $\sigma$ inside the bubble take the form
\begin{align}
&
\begin{cases}
\rho_{-}(\xi)=\sqrt{\frac{2}{\kappa_{5}\epsilon}\biggl[ \sqrt{-9H^4+\kappa_{5}\epsilon\beta_{-}} \, \sinh{\biggl( \sqrt{\frac{2\kappa_{5}\epsilon}{3}}\xi+\xi_{0}^{-} \biggr)} -3H^2 \biggr]}
\\
\\
\sigma_{-}(\xi)=\gamma_{-} \sqrt{\frac{\kappa_{5}\epsilon}{6}} \,\, \frac{\sqrt{-9H^4+\kappa_{5}\epsilon\beta_{-}} \, \cosh{\biggl( \sqrt{\frac{2\kappa_{5}\epsilon}{3}}\xi+\xi_{0}^{-} \biggr)}}{\sqrt{-9H^4+\kappa_{5}\epsilon\beta_{-}} \, \sinh{\biggl( \sqrt{\frac{2\kappa_{5}\epsilon}{3}}\xi+\xi_{0}^{-} \biggr)} -3H^2}
\end{cases}
\text{for} \qquad 0\leq\xi\leq\bar{\xi}
\notag
\end{align}
obtained from \eqref{Case4-general-solution} by requiring that $\rho(0)=0$. This fixes $\xi_{0}^{-}=\arcsinh{\Bigl(\frac{3H^2}{\sqrt{-9H^2+\kappa_{5}\epsilon\beta_{-}}}\Bigr)}$, and $\beta_{-}$ is understood to be positive and $\beta_{-}  > \frac{9H^4}{\kappa_{5}\epsilon}$, which makes $\xi_{0}^{-}$ well defined. Outside the bubble the solution is again \eqref{Vacuum-solution-V=0} and $\beta_{+}\le 0$. With no need to proceed further it is clear from the form of $\rho_{-}(\xi)$ that $\dot\rho_{-}(0)$ diverges which
forces us to discard this case.

\vspace{5mm}
\noindent
\textbf{AdS3.} \qquad The functions $\rho$ and $\sigma$ inside the bubble take now the form
\begin{align}
&
\begin{cases}
\rho_{-}(\xi)=\sqrt{e^{\sqrt{\frac{2\kappa_{5}|V_{c}|}{3}}\xi+\xi_{0}^{-}}-\frac{6H^2}{\kappa_{5}|V_{c}|}}
\\
\\
\sigma_{-}(\xi)=\frac{\gamma\sqrt{\frac{2\kappa_{5}|V_{c}|}{3}}e^{\sqrt{\frac{2\kappa_{5}|V_{c}|}{3}}\xi+\xi_{0}^{-}}}{2\sqrt{e^{\sqrt{\frac{2\kappa_{5}|V_{c}|}{3}}\xi+\xi_{0}^{-}}-\frac{6H^2}{\kappa_{5}|V_{c}|}}}
\end{cases}
\qquad \qquad \qquad  \text{for} \qquad 0\leq\xi\leq\bar{\xi}
\notag
\end{align}
obtained from \eqref{sol.AdS3} by requiring that $\rho(0)=0$, which fixes $\xi_{0}^{-}=\log{\Bigl( \frac{6H^2}{\kappa_{5}|V_{c}|} \Bigr)}$, while outside the bubble the solution is again \eqref{Vacuum-solution-V=0} and $\beta_{+}\le 0$. Like for \textbf{AdS2}, we do not need to proceed further since
$\dot\rho_{-}(0)$ diverges and thus this solution is also discarded.

\section{Conclusions}

In this work, we have computed the false vacuum decay in the presence of gravity and one compact dimension, from dS to Minkowski and from Minkowski to AdS vacua, within the thin wall approximation. The wall can then be described as a $d-2$ brane localised along the radial coordinate of a $d$-dimensional Euclidean brane-world where Israel matching conditions should be imposed on the derivative of the metric parametrising the instanton configuration which drives the transition, across the wall that separates the false and the true vacuum. The discontinuity of the derivatives are then fixed by the brane tension. We have shown that in the non-compact case, this determines the solution and reproduces the same result with the minimisation of the action of the Bounce within the Coleman-de Luccia approach in arbitrary $d$ dimensions.

In the presence of a compact dimension, the main complication arises from the symmetry of the solution which cannot be $O(d)$ but $O(d-1)\times U(1)$ where the abelian factor corresponds to the translation along the compact coordinate enforcing the solution to depend only on one variable, the radial coordinate of $O(d-1)$. Its geometry corresponds to a bubble of a $(d-2)$-dimensional sphere times a circle $S^{d-2}\times S^1$, so that both the radius of the sphere and the radius of the circle depend on the radial coordinate of the bubble $\xi$. The solution is therefore characterised by two functions of the radial coordinate, the radius of the sphere $\rho(\xi)$ and the radius of the circle $\rho\sigma(\xi)$. The Israel matching conditions across the wall, together with the vanishing of the radius of the sphere at the origin of the bubble, determines again the solution uniquely, whose finite action allows to compute the dominant contribution to the transition probability. We provide explicit expressions for the 5-dimensional case, but they can easily be generalised to any dimension $d$ and with any number of compact dimensions.

We found that the exponent of the decay width is given by the tree level result (without gravity) times a gravitational correction depending on one dimensionless parameter $x=\kappa_5 S_1^2/|\epsilon|$, where $S_1/2$ is the brane tension and $\epsilon$ is the energy of the false (true) vacuum in the dS to Minkowski (Minkowski to AdS) transition. This parameter should be less than unity for the effective field theory of Einstein gravity to be valid. It turns out that for the dS to Minkowski transition the correction is a monotonically decreasing function from unity towards zero where the validity of the effective field theory breaks down. On the other hand, for the Minkowski to AdS transition the correction increases from the tree level result which is the same as in the dS to Minkowski case, since in the absence of gravity the absolute energy normalisation is arbitrary.
During the transition, the radius of the extra dimension remains unchanged, up to gravitational corrections.

We expect our analysis to be valid in the region where the size of the compact dimension $R_0$ is less than the radius of the 3-sphere $\bar\rho/H\sim S_1/\epsilon$, a value at which the 5-dimensional non compact result for the decay probability is reproduced, while the $R_0\to 0$ limit leads trivially to the 4-dimensional result, upon appropriate rescaling of the 5-dimensional quantities $S_1$ and $\epsilon$ according to standard dimensional reduction. In the opposite limit, one would expect the CdL $O(5)$ non-compact instanton to be recovered, but the nature and origin of such a transition (if it occurs) is not clear to us and requires a dedicated analysis.   A somewhat similar situation arises in the black hole transition from four to five dimensions where again a maximally spherically symmetric solution does not exist in the presence of a compact dimension but is replaced by a black string which develops a Gregory-Laflamme instability~\cite{Gregory:1993vy}. On the other hand, a different situation arises in de Sitter solutions where the size of a compact dimension is forced to be less than the cosmological horizon by unitarity~\cite{Higuchi:1986py} and there is no transition from dS$_5$ to dS$_4\times S^1$. Actually, by extrapolating our solution to a Lorenzian metric, the same requirement $\rho\sigma<\rho/\dot\rho$ evaluated on the wall leads in the gravity decoupled limit to the bound $R_0\lesssim S_1/\epsilon$, up to a numerical constant. 
For large values of the radius, we expect an instability to show up, invalidating the $O(4)\times U(1)$ solution that should be replaced by the $O(5)$ CdL instanton in order to describe the false vacuum transition. A study of the stability of our solution and its possible breakdown at a critical value of the radius is currently under investigation.

Another interesting question of going beyond the thin wall approximation remains open, in relation to the question of when it breaks down and when the Hawking-Moss classical transition starts dominating.

\acknowledgments
IA is supported by the Second Century Fund (C2F), Chulalongkorn University. The work of DB and HI has been supported by Thailand NSRF via PMU-B, grant number B13F670063. IA would like to thank Matt Kleban for enlightening correspondence.

\bibliography{Bibliography.bib}

\end{document}